 \documentclass[preprint,11pt]{aastex}
 
\usepackage{graphicx}
\usepackage{color}

\newcommand{\etal}{\mbox{et~al.}}

\def\deg      {{\ifmmode^\circ\else$^\circ$\fi}} 

\def\FeII{Fe\,II}
\def\lMgII{Mg\,II$\lambda$2798$\AA$}
\def\MgII{Mg\,II}
\def\km{{\rm\thinspace km\,}} 
\def\kmps{\hbox{\km\ s$^{-1}$\,}} 

 \shorttitle{Evolution of scaling relations for broad line AGN in COSMOS}
 \shortauthors{Merloni et al.}
 
\begin{document}
 
 
 \title{On the cosmic evolution of the scaling relations between black holes and their
   host galaxies: Broad Line AGN in the zCOSMOS survey\altaffilmark{0}} 
 

%
%
%
 \author{
A. Merloni\altaffilmark{1,2},
A. Bongiorno\altaffilmark{2,3},
M. Bolzonella\altaffilmark{4},
M. Brusa\altaffilmark{2},
F. Civano\altaffilmark{5},
A. Comastri\altaffilmark{4},
M. Elvis\altaffilmark{5},
F. Fiore\altaffilmark{6},
R. Gilli\altaffilmark{4},
H. Hao\altaffilmark{5},
K. Jahnke\altaffilmark{7},
A. M. Koekemoer\altaffilmark{8}, 
E. Lusso\altaffilmark{4},
V. Mainieri\altaffilmark{9},
M. Mignoli\altaffilmark{4},
T. Miyaji\altaffilmark{10,11},
A.~Renzini \altaffilmark{12}
M. Salvato\altaffilmark{13,14},
J. Silverman\altaffilmark{15},
J. Trump\altaffilmark{16},
C. Vignali\altaffilmark{17},
G. Zamorani\altaffilmark{4},
P.~Capak \altaffilmark{13,18},
S.~J. Lilly  \altaffilmark{15},
D. Sanders\altaffilmark{19},
Y. Taniguchi\altaffilmark{20},
S.~Bardelli \altaffilmark{4},
C. M. Carollo \altaffilmark{15},
K.~Caputi \altaffilmark{15},
T.~Contini \altaffilmark{21},
G.~Coppa \altaffilmark{4,17},
O.~Cucciati \altaffilmark{22},
S.~de la Torre \altaffilmark{22,23,24},
L.~de Ravel  \altaffilmark{22},
P.~Franzetti \altaffilmark{23},
B.~Garilli \altaffilmark{23},
G. Hasinger\altaffilmark{2,14},
C. Impey\altaffilmark{16},
A.~Iovino \altaffilmark{23},
K. Iwasawa \altaffilmark{4},
P.~Kampczyk \altaffilmark{15},
J.-P. Kneib \altaffilmark{22},
C.~Knobel \altaffilmark{15},
K.~Kova\v{c} \altaffilmark{15},
F.~Lamareille \altaffilmark{21},
J.~-F.~Le Borgne \altaffilmark{21},
V.~Le Brun \altaffilmark{22},
O.~Le F\`evre  \altaffilmark{22},
C.~Maier \altaffilmark{15},
R.~Pello \altaffilmark{21},
Y.~Peng \altaffilmark{15},
E.~Perez Montero \altaffilmark{21,25},
E.~Ricciardelli \altaffilmark{26},
M.~Scodeggio \altaffilmark{23},
M.~Tanaka \altaffilmark{9},
L. A. M. Tasca\altaffilmark{22,23},
L.~Tresse  \altaffilmark{22},
D.~Vergani \altaffilmark{4},
E.~Zucca \altaffilmark{4}
}


 \begin{abstract}
We report on the measurement of the
physical properties (rest frame K-band luminosity and total stellar
mass) of the hosts of 89 broad line (type--1) Active Galactic Nuclei (AGN)  detected
in the zCOSMOS survey in the
redshift range $1<z<2.2$. The unprecedented multi-wavelength coverage of
the survey field allows us to
disentangle the emission of the host galaxy from that of the nuclear
black hole in their Spectral Energy Distributions (SED).
We derive an estimate of black hole masses
through the analysis of the broad \MgII\ emission
lines observed in the
medium-resolution spectra taken with {\it VIMOS/VLT} as part of the
zCOSMOS project. 
We found that, as compared to the local value, the
average black hole to host galaxy mass ratio appears to evolve positively with
redshift, with a best fit evolution of the form
$(1+z)^{0.68 \pm 0.12 ^{+0.6}_{-0.3}}$, where the large asymmetric systematic
errors stem from the uncertainties in the choice of IMF, in the
calibration of the virial relation used to estimate BH masses and in
the mean QSO SED adopted. On the other hand, if we consider the
observed rest frame K-band luminosity, objects tend to be brighter,
for a given black hole mass, than those on the local $M_{\rm
  BH}$-$M_K$ relation. This fact, together with more indirect evidence
from the SED fitting itself, suggests that the AGN hosts are likely actively
star forming galaxies.  
A thorough analysis of observational biases induced by
intrinsic scatter in the scaling relations reinforces the conclusion
that an evolution of the $M_{\rm BH}-M_{*}$ relation must
ensue for actively growing black holes at early times:
either its overall normalization, or its intrinsic scatter (or
both) appear to increase with redshift. 
This can be interpreted as signature of either a more rapid growth of supermassive
black holes at high redshift, a change of structural properties of AGN
hosts at earlier times, or a significant mismatch between the typical growth
times of nuclear black holes and host galaxies. In any case, our results
provide important clues on the nature of the early co-evolution of
black holes and galaxies and challenging tests
for models of AGN feedback and self-regulated growth of structures. 
\end{abstract}

 
\altaffiltext{0}{Based on observations obtained at the European Southern Observatory 
(ESO) Very Large Telescope (VLT), Paranal, Chile, as part of the Large 
Program 175.A-0839 (the zCOSMOS Spectroscopic Redshift Survey). Also
based on observations with the NASA/ESA {\em
Hubble Space Telescope}, obtained at the Space Telescope Science
Institute, which is operated by AURA Inc, under NASA contract NAS
5-26555; and on data collected at: the Subaru Telescope, which is operated by
the National Astronomical Observatory of Japan; the XMM-Newton, an ESA science mission with
instruments and contributions directly funded by ESA Member States and
NASA; the European Southern Observatory under Large Program
175.A-0839, Chile;  the National Radio Astronomy Observatory which is a facility of the National Science 
Foundation operated under cooperative agreement by Associated Universities, Inc.;
and the Canada-France-Hawaii Telescope with MegaPrime/MegaCam operated as a
joint project by the CFHT Corporation, CEA/DAPNIA, the National Research
Council of Canada, the Canadian Astronomy Data Centre, the Centre National
de la Recherche Scientifique de France, TERAPIX and the University of
Hawaii.}  
\altaffiltext{1}{Excellence Cluster Universe, TUM, Boltzmannstr. 2,
  85748, Garching, Germany}
\altaffiltext{2}{Max Planck Institut f\"ur Extraterrestrische Physik,
  Giessenbachstr., 85471 Garching, Germany}
\altaffiltext{3}{University of Maryland, Baltimore County, 1000 Hilltop Circle, Baltimore, MD21250, USA}
\altaffiltext{4}{INAF-Osservatorio Astronomico di Bologna, via Ranzani 1, I-40127 Bologna, Italy}
\altaffiltext{5}{Harvard-Smithsonian Center for Astrophysics, 60 Garden Street, Cambridge, MA 02138}
\altaffiltext{6}{INAF-Osservatorio Astronomico di Roma, via Frascati 33,
Monteporzio (Rm), I00040, Italy}
\altaffiltext{7}{Max Planck Institut f\"ur Astronomie, K\"onigstuhl 17, Heidelberg, D-69117, Germany}
\altaffiltext{8}{Space Telescope Science Institute, 3700 San Martin
Drive, Baltimore, MD 21218, USA} 

\altaffiltext{9}{ESO, Karl-Schwarzschild-Strasse 2, D-85748 Garching, Germany}
\altaffiltext{10}{Instituto de Astronomia, UNAM-Ensenada, Km 103 Carretera Tijuana-Ensenada, 22860 Ensenada, BC Mexico}
\altaffiltext{11}{Center for Astrophysics and Space Sciences,
University of California at San Diego, Code 0424, 9500 Gilman Drive,
La Jolla, CA 92093, USA}

\altaffiltext{12}{INAF - Osservatorio Astronomico di Padova, Padova, Italy}

\altaffiltext{13}{California Institute of Technology, MC 105-24, 1200
East California Boulevard, Pasadena, CA 91125, USA}
\altaffiltext{14}{Max-Planck-Institute f\"ur Plasmaphysik,
  Boltzmannstrasse 2, D-85748 Garching, Germany}
\altaffiltext{15}{Department of Physics, ETH Zurich, CH-8093 Zurich, Switzerland}
\altaffiltext{16}{Steward Observatory, University of Arizona, 933 North Cherry Avenue, Tucson, AZ 85721, USA}
\altaffiltext{17}{Dipartimento di Astronomia, Universit{\'a} di Bologna, via Ranzani 1, I-40127, Bologna, Italy}
\altaffiltext{18}{Spitzer Science Center, 314-6 Caltech, Pasadena, CA 91125, USA}
\altaffiltext{19}{Institute for Astronomy, University of Hawaii, 2680 Woodlawn Drive, HI 96822, USA}
\altaffiltext{20}{Research Centre for Space and Cosmic Evolution, Ehime University, Bunkyo-cho 2-5, Matsuyama 790-8577, Japan}
\altaffiltext{21}{Laboratoire d'Astrophysique de Toulouse-Tarbes, Universit{\'e} de Toulouse, CNRS, 14 avenue Edouard Belin, F-31400 Toulouse, France}
\altaffiltext{22}{Laboratoire d'Astrophysique de Marseille, 
   CNRS-Univerist{\'e} d'Aix-Marseille, 38 rue Frederic Joliot Curie, 
   13388 Marseille Cedex 13, France}
\altaffiltext{23}{INAF-IASF, Via Bassini 15, I-20133, Milano, Italy}
\altaffiltext{24}{INAF Osservatorio Astronomico di Brera, Via Brera 28, I-20121 Milano, Italy}
\altaffiltext{25}{Instituto de Astrofisica de Andalucia, CSIC, Apdo. 3004, 18080, Granada, Spain}
\altaffiltext{26}{Dipartimento di Astronomia, Universit{\'a} di Padova, vicolo Osservatorio 3, I-35122 Padova, Italy}


 
 

 
\section{Introduction}
Tight scaling relations between the central black holes mass and
various properties of their host spheroids (velocity dispersion, $\sigma_{*}$, stellar
mass, $M_{*}$, luminosity, core mass deficit) characterize the structure of nearby inactive galaxies
\citep{magorrian:98,gebhardt:00,ferrarese:00,tremaine:02,marconi:03,haering:04,hopkins:07b,kormendy:09,gultekin:09}.
A result of the search for local QSO relics via the study of their
dynamical influence on the surrounding stars and gas made possible by
the launch of the {\it Hubble Space Telescope} nearly twenty years
ago, these correlations have revolutionized the way we conceive the
physical link between galaxy and AGN evolution.
Coupled with the fact that supermassive black holes (SMBH) growth is
now known to be due mainly to
radiatively efficient accretion over cosmological times, taking place
during ``active'' phases
\citep{soltan:82,marconi:04,shankar:04,merloni:08}, this led to the
suggestion that most, if not all, galaxies went through a phase of
 nuclear activity in the past, during which a strong physical
coupling (generally termed 'feedback') must have established a
long-lasting link between host's and black hole's properties.
 
This shift of paradigm has sparked the activity of theoretical
modelers. Following  pioneering analytic works (Ciotti \&
Ostriker 1997,2001; Silk \& Rees 1998;
Fabian 1999; Cavaliere \& Vittorini 2002; Whyithe \& Loeb 2003), widely
different approaches have been taken to study the role of AGN in
galaxy evolution. Semi-analytic models (SAM) have been the most
numerous (Monaco et al. 2000; Kauffmann \& Haehnelt 2000; Volonteri et
al. 2003; Granato et al. 2004; Menci et al. 2006; Croton et al. 2006;
Bower et al. 2006; Malbon et al. 2007; Marulli et al. 2008; Somerville
et al. 2008), and, among other things, have helped
establishing the importance of late-time feedback from radio-active
AGN for the high-mass end of the galaxy mass function. However, firmer
conclusions on the physical nature of such a feedback mode have been
hampered by the large freedom SAM have in choosing baryonic physics
recipes to implement in their schemes. On the other hand, fully
hydrodynamic simulations of the cosmological evolution of SMBH have
been also performed (Di Matteo et al. 2003, 2008; Sijacki et
al. 2007; Coldberg \& Di Matteo 2008), but their computational costs have so far allowed only a
limited exploration of sub-grid prescriptions (AGN physical models) in
relatively small cosmological volumes. A
third, hybrid, approach has also been followed, in which the results
of high-resolution hydrodynamic simulations of galaxy-galaxy mergers
with black holes (Di Matteo et al. 2005; Springel et al. 2005) have
been used to construct a general framework for merger-induced AGN
feedback, capable of passing numerous observational tests (Hopkins et
al. 2006). 

Almost all the approaches outlined above use the local scaling
relations as a constraint to the model parameters.
As such, these relations have proved themselves unable to
unambiguously determine the physical nature of the SMBH-galaxy
coupling. One obvious way out of this impasse is to study their 
redshift evolution, that different models predict
to be different
\citep{granato:04,robertson:06,croton:06b,fontanot:06,malbon:07,marulli:08,hopkins:09a}.

From the observational point of view, the current situation is far
from being clear. In recent years, a number of groups have employed
different techniques to try and detect signs of evolution in any of
the locally observed scaling relations. Most efforts have been devoted
to the study of the $M_{\rm BH}-\sigma_{*}$ relation. Shields et
al. (2003) and Salviander et al. (2007) have used narrow
nebular emission lines ([OIII], [OII]) excited by the AGN emission in the nuclear
region of galaxies as proxies for the central velocity dispersion, and
compared these to the black hole mass estimated from the broad line
width of QSOs from $z\sim 0$ to $z\sim 3$ (see \S \ref{sec:virial}). In both cases, a large
scatter has been found in the relation between $M_{\rm BH}$ and
$\sigma_{*}$, and the results  are either in favor of (Salviander et al. 2007) or
against \citep{shields:03} a positive evolution of the black
hole mass to host dispersion ratio. However, as pointed out by
Botte et al. (2004) and Greene \&  Ho (2005), there are a number of problems with the
underlying assumption that the narrow emission lines are good probes
of the central gravitational potential, and the systematic
uncertainties this method is endowed with are large. Komossa and Xu
(2007) have also shown that the the AGN for which the [OIII] emission
line width is far broader that the host galaxy's stellar velocity
dispersion $\sigma_{*}$ tend to show clear sign of blue-shift in the
narrow emission line (of the order of 100 km/s or larger). Thus, using [OIII] as a proxy for $\sigma_{*}$
would at least require good enough spectral resolution to measure such blue-shifts.

An alternative path has been followed by Woo et al. (2006), Treu et
al. (2007) and Woo et al. (2008), who have studied carefully selected
samples of moderately bright AGN in narrow redshift ranges ($z\sim
0.36$ and $0.57$), where the host's stellar velocity dispersion can be measured
directly from the absorption lines in high signal-to-noise spectra. They 
also found evidence of (strong) positive evolution of the $M_{\rm BH}$ to
$\sigma_{*}$ ratio compared to the local value. 
This method, although promising and reliable, is quite
inefficient and telescope-time consuming: secure detection of spectral
absorption features in massive ellipticals at $1\la z \la 2$ require
hundreds of hours of integration time on a 8-meter class telescope \citep{cimatti:08}.

Other groups have chosen to try and derive information on the host
mass of broad line AGN using multi-colour 
image decomposition techniques \citep{jahnke:04,sanchez:04}
or spatially deconvolving optical spectra \citep{letawe:07}. Due to the
severe surface brightness dimming effects, employing these techniques
for redshift QSOs becomes increasingly challenging (but see Schramm,
Wisotzki and Jahnke 2008; Jahnke et al. 2009; Decarli et al. 2009, in prep.),
unless gravitationally lensed QSOs are selected \citep{peng:06b,ross:09}. In
all cases, very deep, high resolution optical images ({\it HST}) are
necessary to reliably disentangle the nuclear from the host galaxy
emission. Statistically, the most significant results have been
published by Peng et al. (2006b), who, based on a large sample of
51 AGN (both lensed and nonlensed) in the range $1<z<4.5$, observed that
the ratio $M_{\rm BH}/M_{*}$ increases with look-back time, up to a
factor $\simeq 4^{+2}_{-1}$ at the highest redshift probed.  

Also cold/molecular gas motions on large galactic scales have been
used to infer the properties (total mass in particular) of the hosts
of AGN at different redshifts. In the local Universe, HI 21 cm lines
have been used by Ho et al. (2008) to derive total stellar masses of
the bulges around Seyfert-like AGN. 
Using instead CO lines to estimate the velocity dispersion in
high-redshift QSO hosts, Walter et al. (2004) and Shields et
al. (2006) find tentative evidence
that the hosts are very under-massive compared to their central BHs (i.e.,
a positive redshift evolution of the  $M_{\rm BH}/M_{*}$ ratio), a
result confirmed by the study of Ho (2007).

Finally, a completely different approach has been that of trying to follow and compare the
evolution of global descriptors of the galaxy and SMBH populations,
such as mass densities (Merloni, Rudnick and Di Matteo 2004; Hopkins
et al. 2006b; Shankar et al. 2008). Using the simple {\it ansatz} that total
black hole mass density can only increase with time, and requiring
that the limits imposed by local demographics are not violated, 
these works showed in general
very moderate, if any, signs of cosmological evolution of the average black
hole to host mass ratio.

Here we present a new method to tackle the issue of studying black
hole-galaxy scaling relations at high redshift. Starting from a sample
of un-obscured AGN, for which the broad line kinematics can be used to
infer the central SMBH mass, we take advantage of the unprecedentedly
deep multi-wavelength coverage of the COSMOS \citep{scoville:07} field
and develop a novel SED fitting technique that allows us to decompose
the entire spectral energy distribution into a nuclear AGN and a host
galaxy components \citep{bongiorno:07}. We show here how, for the majority of the objects
in our sample, rest frame K-band luminosity and total stellar
mass of the host can be robustly determined, opening the way to a
detailed study of the scaling relations in type--1 AGN at $1\la z \la
2.2$.

The structure of the paper is the following: we will begin (\S~\ref{sec:sample})
by introducing our sample, before proceeding to a discussion of our
SED fitting method in section~\ref{sec:sed_fit}, focusing our
attention on the measures of the rest-frame K-band luminosity of the
AGN hosts (\S~\ref{sec:mk}), on their total stellar mass (\S~\ref{sec:mstar}). In section~\ref{sec:virial}
we will describe our estimates of the black hole masses obtained by studying the
properties of the broad \MgII\ emission line, while in
\S~\ref{sec:bhmass} we will briefly outline the characteristics of our
AGN sample in terms of bolometric luminosity, BH mass and accretion
rates. Section~\ref{sec:scaling_evol} contains the main novel results
of our study, namely the analysis of the scaling relation for the
objects in our sample, as well as a characterization of their observed
redshift evolution. An important part of our analysis, however, is the
assessment of possible observational biases responsible for the
trend observed, that we carry out in section~\ref{sec:bias}. This
allows us to reach robust conclusions, that we discuss at the end of
the paper, in section~\ref{sec:disc}. 

Throughout this paper, we use the standard cosmology ($\Omega_m=0.3$,
$\Omega_{\Lambda}=0.7$, with $H_0=$70 km s$^{-1}$ Mpc$^{-1}$).

\section{The zCOSMOS type--1 AGN sample}
\label{sec:sample}
Our AGN sample consists of the sub-sample of
objects in the zCOSMOS bright spectroscopic catalog \citep{lilly:07} for 
which one or more broad emission lines have been identified in the
spectrum. As such, it will be in the following identified as either
a Broad Line AGN (BLAGN) or a type--1 AGN sample, without introducing
any distinction between the two terms. 

The zCOSMOS bright sample consists, at the times of writing, of 10,644 (medium
resolution, MR) spectra observed 
 with the {\it VIMOS} multi-object spectrograph on ESO-VLT
 in the COSMOS field, selected only on the basis 
of their I$_{AB}$ magnitude (I$_{AB}<$22.5), based on the {\it
  HST}/ACS imaging of the COSMOS field \citep{koekemoer:07}.
For a detailed description of the spectroscopic survey we
refer the reader to \cite{lilly:07,lilly:09}.

Within the zCOSMOS database, we have selected objects with broad
emission lines full width half maximum (FWHM)
larger than $2000$ ${\rm km\,s^{-1}}$, a secure threshold for truly
broadened lines, as compared to our
spectral resolution (R $\sim$ 580 for the MR grism, corresponding to
$\sim 4.8\AA$ and $\sim 520$ km/s at the wavelength of \MgII\ emission).   
The final sample of type--1 AGN spectra selected from the zCOSMOS survey
consists of 164 objects which correspond to about 1.8\% 
of the objects in the total 10k zCOSMOS database with measured redshift.

We measure black hole masses by applying the ``virial'' or
``empirically calibrated photo-ionization'' method
\citep{wandel:99,kaspi:00,peterson:04,bentz:06} to the \lMgII\ broad emission
line. We have selected BLAGN in the redshift interval z
$\sim$ [1.06, 2.19], within which \lMgII\ can be measured
reliably. This takes into  account the expected broad line width and the problems occurring 
at the edge of the spectrograph (e.g. fringing in the red part of the spectrum)
we estimate that the spectral range available to measure \MgII\
line widths 
is $\sim$ 5650\AA-9150\AA\
(VIMOS spectral wavelengths range from 5500\AA\ to 9500\AA). 

Within this range, the zCOSMOS bright sample contains 
104 AGN. After a quick inspection, 15 of them have been 
excluded from the analysis because of the low quality of the available spectra, 
leaving us with 89 objects. Ten of those have radio counterparts (at
1.4 GHz) in
the VLA/COSMOS catalogues (Schinnerer et al. 2007; Bondi et al. 2008), and are listed in Table~\ref{tab:agn}.

\subsection{Photometry}
The zCOSMOS BLAGN sample has been cross correlated with
the optical multi-band catalog of Capak et al. (2007), 
the CFHT/K band catalog of Mc Cracken et al. (2009), 
the IRAC  catalog by  Sanders et al. (2007) and the MIPS catalog by
Le Floch et al. (2009).
Briefly, the optical catalog contains about 3 million objects
detected in at least one of the Subaru bands (b,v,g,r,i,z) down to a AB
magnitude limit of $\sim 27$ (see Capak et al. 2007, Taniguchi et al. 2007 for
more details). From this catalog is possible to extract a subsample of
$\sim 1.3$ million sources which have signal to noise $>5$ in the i- or
z-bands.
The K-band catalog contains about 5$\times10^5$ galaxies detected
at a S/N$>$5 down to K(AB)=23.5 (Mc Cracken et al. 2009). 
The IRAC catalog contains about 4$\times10^5$
objects detected in the 3.6 micron (IRAC channel 1) band and it is 90\%
complete at $>1 \mu$Jy (AB=23.9). For each source in the catalog,
the photometry from all the other IRAC channels is also reported. 
The MIPS catalog, obtained in Cycles 2, 3 and 4, has very accurate photometry
(Sanders et al. 2007). 
As described in details in Salvato et al. (2009), 
the fluxes in the optical and NIR bands were 
measured in fixed apertures of 3" diameter, on point-spread 
function (PSF) matched images (FWHM of 1.''5). Monte-Carlo simulations 
(Capak et al. 2007) have been used to correct for the flux
potentially missed within the apertures. 
The 3.''8 aperture fluxes given in the the COSMOS-IRAC 
catalog sources  were also converted to total 
fluxes by using conversion  factors taken from Surace et al. (2005).

To make sure that confusion is not an issue in
the IRAC and MIPS bands, 
we have visually inspected all IRAC and MIPS matches. In the IRAC
bands, 82/89 objects have secure counterparts, and for them
no obvious case of blending has been found when looking into 3"
diameter circles centered on the source of the optical photometry: only
one IRAC source is found in all objects, and contaminating flux is
usually lower than 1\%. 
For MIPS sources, the situation is slightly different.
There we found 4 cases where the MIPS
source could indeed be a blend of two IRAC sources lying within the
MIPS error circle. We have thus decided to remove the MIPS photometric
point for these objects. In any case, we have verified that our
results are not changed, neither quantitatively nor qualitatively if
we recalculate the stellar masses increasing the error uncertainty on
all the 24$\mu$m points by $\pm$ 50\%.

In summary, apart from the 7 cases with ambiguous identification
of the IRAC counterparts and the 4 confused MIPS sources, 
for all remaining 78 objects in our final sample, we have used 14
different bands that encompass optical to MIR wavelengths: 
6 SUBARU bands (B, V, g, r, i, z); U, J and K band
from CFHT + 4 Spitzer/IRAC 
bands + 24$\mu$m form Spitzer/MIPS. This allows us
to sample a wide wavelength interval, ranging from $\sim$3800 \AA
 (U$_{\rm CFHT}$) to 24$\mu$m. All errors quoted are Poissonian.

\section{Disentangling the AGN and host galaxy emission with SED fitting}
\label{sec:sed_fit}
One of the crucial goals of our study is to use the unprecedented 
multi-wavelength coverage of the COSMOS field to robustly derive
host-galaxy properties through detailed model fitting of the total SED
of the broad line AGN in our sample. Salvato et al. (2009) have
demonstrated how the COSMOS data allow the 
determination of reliable photometric redshifts by using composite
AGN+galaxy templates to fit the multi-band photometry of all XMM-COSMOS
sources (including both obscured and un-obscured AGN).
Here we apply a similar technique to our sample of BLAGN (with known
spectroscopic redshift) to try to unveil the physical properties of
the galaxy component. 
We fit the observed SED with a relatively large grid of models made
from a combination of AGN and host galaxy
templates. For the AGN component we adopt the Richards et al.~(2006)
mean QSO SED (but see \S~\ref{sec:agn_template_heng} for a discussion
of possible alternative choices), as derived from the study of 259 IR selected quasars with both Sloan
Digital Sky Survey and Spitzer photometry. 
We allow for extinction of the nuclear AGN light 
applying a SMC-like dust-reddening law \citep{prevot:84} of the form:
$A_{\lambda}/E(B-V) = 1.39 \ \lambda_{\mu m}^{-1.2}$
for $E(B-V)$ (reddening factor) values in the range $0\le E(B-V) \le
0.3$. 
For the host galaxy component, we adopt two different sets of templates:

\begin{itemize}
\item[i)] First of all, we use the library of (observed) galaxy templates produced by the
SWIRE survey  (see Polletta et al. 2007, hereafter P07). From the
entire library available of 25 templates, we 
excluded the AGN and composites (starburst+AGN), thus retaining only
14 templates (3 Ellipticals, 7 Spirals  
and 4 Starburst). Such a fitting algorithm has four free parameters: two normalizations for the AGN and galaxy templates, respectively,
and two corresponding reddening factors.
\item[ii)] We also created our own library of synthetic spectra using the
well known models of stellar population synthesis of Bruzual and
Charlot (2003, hereafter BC03). Similarly to previous studies of the
galaxy population in COSMOS (Ilbert et al. 2009; Bolzonella et al. 2009), 
we built 10 exponentially
declining star formation histories (SFH) $SFR\propto e^{-t_{\rm age}/\tau}$ with
e-folding times, $\tau$, ranging
from 0.1 to 30 Gyr, plus a model with constant star formation. For
each of these SFH, we calculate the synthetic spectrum at different
ages, $t_{\rm age}$, ranging from 50 Myr to 5 Gyr, subject only to the constraint that
the age should be smaller than the age of the Universe at the redshift
of the source. Finally, we allow for dust extinction, modeled by means
of the Calzetti's law (Calzetti et al. 2000), with values in the range
$0\le E(B-V) \le 0.5$.  Following Fontana et al. (2006)  and Pozzetti
et al. (2007), we impose the prior $E(B-V) < 0.15$ if $t_{\rm
  age}/\tau > 4$ (a significant extinction is only allowed for 
galaxies with a high SFR). We adopt a Salpeter (1955) Initial Mass Function (IMF) to compute stellar
masses. Different choices of IMF lead to systematic shifts in the
estimated stellar masses for any given SED (with the maximum shift for
a Chabrier IMF \citep{chabrier:02} given by $M_{\rm *,Chabrier}\approx M_{\rm
  *,Salpeter}/1.8$; 
see e.g. Pozzetti et
al. 2007; Ilbert et al. 2009). We will discuss the effect of the
IMF on our results in section~\ref{sec:bias}. Such a fitting algorithm
has six free parameters: two normalizations for the AGN and galaxy
templates, respectively, two corresponding reddening factors, and the
age and e-folding time of the exponentially declining SFH.
\end{itemize}

\begin{figure}
\includegraphics[scale=0.25,angle=270]{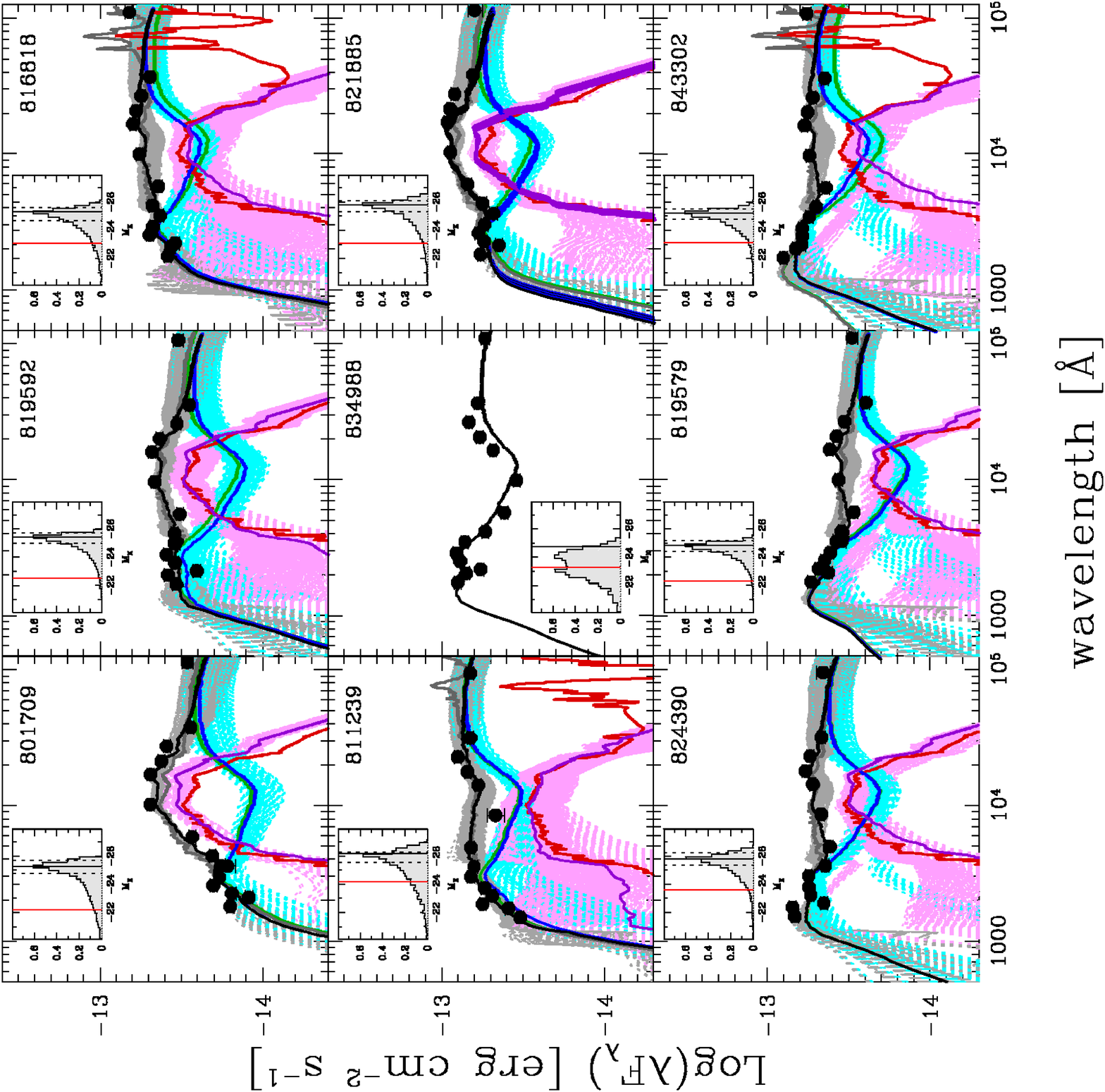}
\caption{Examples of SED decompositions. Black circles are rest-frame
  fluxes corresponding to the 14 bands used to constrain the SED of
  each object.   
Purple and blue lines correspond respectively to the galaxy and the AGN
template found as best fit solution through the $\chi^2$ minimization
for the BC03 template set (red and dark green for the P07 one), while the 
black line shows their sum (dark grey for P07 total). 
Pink and cyan dotted lines show the
 range of allowed SED from the BC03 template library within 1$\sigma$
 of the best-fit $M_K$ measure, and light gray their sum. 
For one objects (VIMOS IDs 834988)  our fitting
procedure returns only an upper limit for the galaxy rest-frame K-band
magnitude (overall 10/89); in this case we only plot the AGN spectral
component. The inset in each panel shows the normalized probability
distribution, $P=\exp{(-\chi_{\rm red}^2/2)}$, for the rest-frame K-band
absolute magnitude of the host galaxy, with the solid vertical
line marking the best fit value and the dashed lines the 1-$\sigma$
uncertainties. The red vertical line marks instead 5\% of the K-band
magnitude of the AGN component. 
The full set of images of SED decomposition
can be found at this URL: {\tt
  http://www.mpe.mpg.de/$\sim$am/plot\_sed\_all\_rev.pdf}} 
\label{fig:sed_decompose}
\end{figure}

We examined the global spectral energy distribution of each object and we fit the observed
fluxes using a combination of AGN and galaxy emission using the
templates extracted from our libraries. 
Nine examples of the SED fitting are shown in
Figure~\ref{fig:sed_decompose} (a gallery of all 89
SEDs can be found at {\tt www.mpe.mpg.de/$\sim$am/plot\_sed\_all\_rev.pdf}).

For each parameter of interest ($M_K$, $M_{*}$), 
we compute the one-dimensional
$\chi^2$ distribution obtained marginalizing over all the other
parameters (with flat priors). The normalized probability
distributions of $M_K$ ($P\propto \exp{(-\chi_{\rm red}^2/2)}$) for a few objects
are shown as insets in  Fig.~\ref{fig:sed_decompose}.
As a general rule, because we are fitting the data with the sum of two
model components, the probability distributions 
are asymmetric, with sometimes large tails towards small values of
these parameters, corresponding to the cases in which the fitting
procedure does not require with high significance the galaxy component
besides the AGN one. 
We determine the best fit
value of the parameter of interest as the value that minimizes the
$\chi^2$. One sigma errors on the best fit parameters are computed rescaling the
observational uncertainties until the minimum reduced $\chi_{\rm
  red}^2=1$ (for a number of degrees of freedom equal to 8 for 79
objects, 7 for 4 without MIPS and 5 for the 7 without clear IRAC
counterparts), and then 
finding the range in the parameter values within which $\Delta\chi^2
\le 1$.
We then assign just an upper limit to the host galaxy
rest-frame K-band magnitude in all those cases where, within the
above-mentioned uncertainty  (i.e. where $\Delta\chi^2 \le 1$),
the rest frame K-band luminosity  can have values smaller than a fixed
fraction $f_{\rm gal}$ of the corresponding AGN luminosity in the same
rest-frame band. By inspection of the best fit SED, after a number of
trials, we fix $f_{\rm gal}=0.05$; simply put, if we find a
significant probability that the galaxy component is smaller than 5\%
of the AGN one, we decide that only an upper limit to the galaxy
luminosity can be assigned.
These 5\% limit values are marked as vertical red
lines in the insets of Fig.~\ref{fig:sed_decompose}. Increasing such a
threshold to 0.1, although almost doubling the number of objects
with only upper limits on the host galaxy $M_K$, hardly produce any
significant change in the general results and in the global trends
discussed in section~\ref{sec:scaling_evol}.
In those cases, the value of the upper limit is taken as the (non-zero) value of
the parameter associated to the highest possible value of $M_K$ within
the uncertainty. Finally, whenever we decide that only an upper limit
can be meaningfully associated to the K-band luminosity of the host
galaxy, we assign an upper limit to the object's total stellar mass
adopting the median mass-to-light ratio of all other objects in the
sample (this corresponds to adopting the following relation between
the logarithm of the total stellar mass and $M_K$ for the upper limits
in the sample: ${\rm Log}M_{*}=-0.55-0.4(M_K-3.28)$, see \S~\ref{sec:mstar}).
In total, for 10/89 objects we can provide only upper limits for
the host galaxy SED component (and thus for $M_K$ and $M_{*}$).

\subsection{Rest frame K band luminosities}
\label{sec:mk}
We are mainly interested here in determining the total mass of the
host. It is generally believed that local scaling relations 
apply only when the bulge/spheroid component of the host galaxy is
considered (Kormendy and Gebhardt 2001; but see the recent works of
Kim et al. 2008, Bennert et al. 2009, for a different view); 
however, reliable bulge-disk decomposition for our AGN
hosts are problematic and will not be considered here; we will briefly
discuss the implications of this issue later, in section~\ref{sec:bias_imf}. 
In most of the objects of the sample the
nuclear AGN emission dominates the emission in all optical bands, 
and the constraints on the host galaxy emission are derived mostly
in the wavelength
range where the AGN SED has a minimum, around 1.2 $\mu$m, \citep{elvis:94,richards:06}. 
This is close to the rest frame K-band, which is itself a good
(i. e. such that the mass-to-light ratio in the K band has a
1-$\sigma$ scatter of about 0.1 dex)
indicator of total mass (see e.g. Madau, Pozzetti \& Dickinson 1998,
Bell et al. 2003).
We thus proceed in two steps: we first try and constrain the
rest-frame K band magnitude of the AGN hosts, $M_{K}$, for a comparison
with the local $M_{\rm BH}$-$M_K$ scaling relations (Marconi and
Hunt 2003; Graham 2007). Then, we proceed to  
derive the confidence interval on the measure of the total stellar
mass.

As a first step, we verified that, given our choice of AGN template,
the derived K-band rest frame magnitudes of the host galaxies are not sensitive to the
particular set of galaxy SED templates used (P07 vs. BC03).
 The average 
1-$\sigma$ errors in the estimated $M_K$ depend on the model SED used
to fit the data, and are approximately 0.4 and 0.3 magnitudes for
BC03 and P07, respectively. Keeping this in mind, we find a
good agreement between these two methods: the difference in $M_K$
 between the estimates obtained using the P07
templates and those obtained using the BC03 ones is strongly peaked at
around zero, with small scatter (of the order of 0.3 dex), with the exception of 8/89
outliers (i.e. objects with $|\Delta M_K| >0.7$, larger than the sum of typical
1-$\sigma$ errors in $M_K$). The estimated $M_K$ (in Vega\footnote{Since our magnitudes were all calibrated on the AB
  system, in order to ease the comparison with literature work, we use the
  following conversion between Vega and AB COSMOS K-band magnitudes:
  $M_{K,{\rm Vega}}=M_{K,{\rm AB}}-1.84$}, for the fits with the BC03
templates) are given in Table~\ref{tab:bc}, while the distribution of
the estimated $M_K$ is shown in the upper panel of
Fig.~\ref{fig:mlk}. There, as a term of reference, we have also plotted
the corresponding distribution of the rest-frame K-band magnitudes of
the AGN components from our fits. Typically, AGN are 1-2
magnitudes brighter than their hosts in the K-band; thus, the fraction of K-band light contributed by
the host galaxy has a median of about 25-30\%, with 90\% of the object
having this ratio smaller than 0.4. The ratio $f_{\rm
  gal,K}\equiv\frac{L_{\rm gal}}{L_{\rm AGN}}$ of the galaxy-to-AGN
  luminosity in the rest frame K-band is also given in table~\ref{tab:bc}.

Encouraged by the robustness of the $M_K$ determination for the AGN
hosts in our sample, we proceed to the discussion of the stellar mass
estimates, and refer the reader to section~\ref{sec:agn_template_heng}
for a further discussion of how a different choice of AGN SED template
could modify our results. 

\subsection{Host galaxy masses}
\label{sec:mstar}
\begin{figure}[ht]
\epsscale{1.0}
\plotone{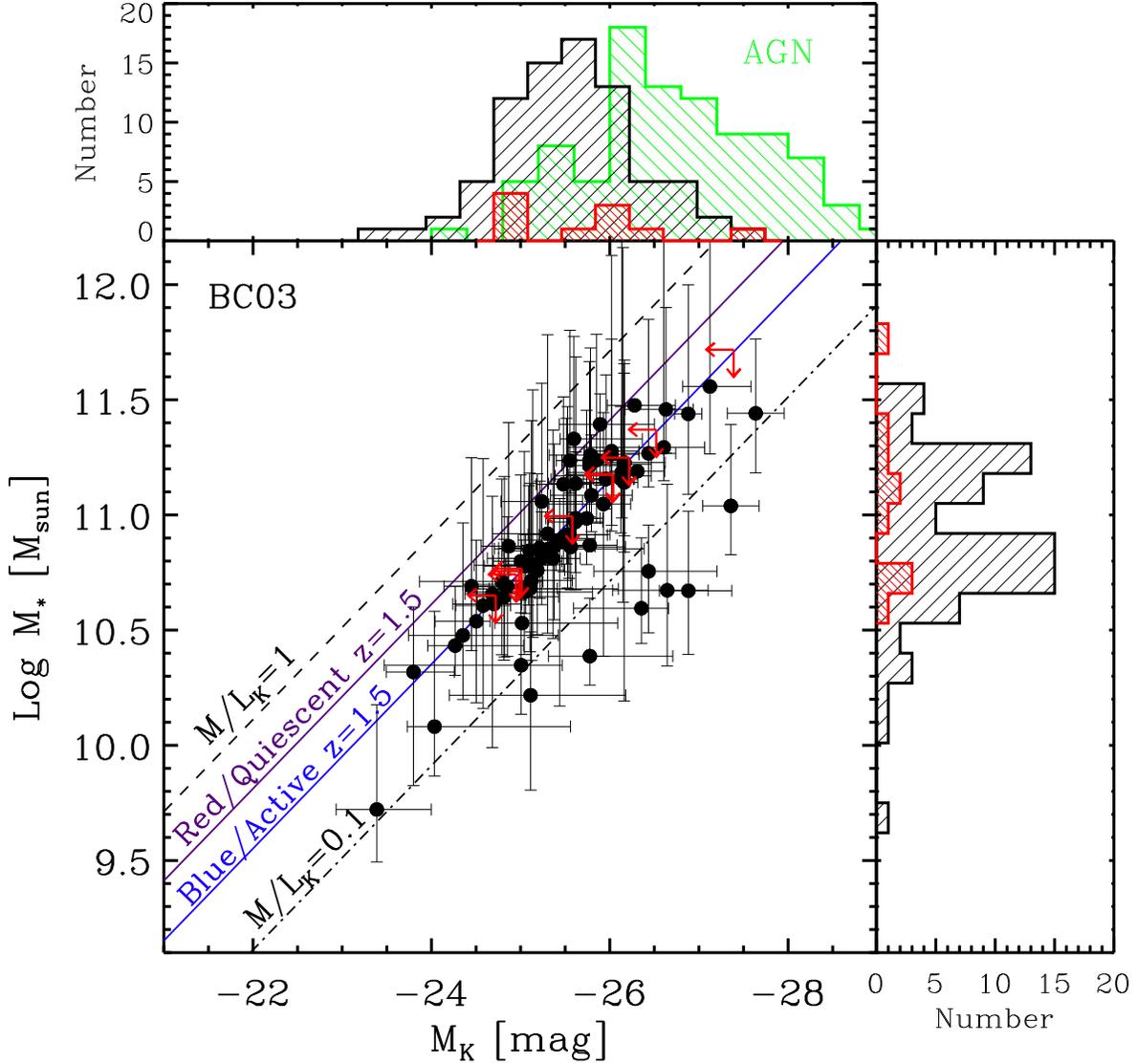}
\caption{Estimated rest-frame K band absolute magnitude vs. total 
stellar mass for the type--1 AGN hosts of our sample (large
panel). Black solid circles are measures, red arrows upper limits. 
As a reference, lines of constant mass to light ratio ($M/L_K$) equal to 1 and 0.1 
are plotted as dashed and dot-dashed lines, respectively. The blue line is
the average mass to light ratio  for blue/actively
star-forming galaxies at $z=1.5$ computed with the Arnouts et
al. (2007) relation and shifted by -0.1 dex to make it consistent with
the S-COSMOS results (Ilbert et al. 2009):
${\rm Log}M/L_{K}=0.27 z - 0.15$. The purple line is
the average mass to light ratio for red/quiescent
galaxies at $z=1.5$ computed with the Arnouts et
al. (2007) relation, also shifted by -0.1 dex (Ilbert et al. 2009):
${\rm Log}M/L_{K}=0.17 z - 0.05$. The upper small panel shows the
distribution of $M_K$ (black histogram: detections, red histogram:
upper limits) together with the distribution of AGN K-band rest frame
magnitudes (green histogram). In the right small panel the
distribution of the measured stellar masses is displayed.} 
\label{fig:mlk}
\end{figure}

The fitting procedure with the BC03 templates allows us to estimate also the total stellar mass
of the AGN hosts: each combination of SFH, $\tau$ and $t_{\rm age}$
(see \S~\ref{sec:sed_fit}) is uniquely associated to a value of SFR
and total stellar mass, computed taking into account the effect of stellar mass
loss. 
The distribution of these stellar mass measurements is shown in the
right panel of Figure~\ref{fig:mlk}. 
The estimated Log$M_{*}$ (for the fits with the BC03
templates) are also
given in Table~\ref{tab:bc}.

What kind of galaxies are these? As we will discuss in more detail
later on, it is extremely difficult to extract reliable information on
the star-forming properties of these objects, due to the dominant
presence of the AGN emission. A more statistical comparison can be
made by comparing the
inferred masses with the overall galaxy mass function in the same
redshift range as obtained by the S-COSMOS survey 
(duly shifted in mass to account for the difference in the
average stellar mass between the Chabrier and Salpeter IMF, see Ilbert
et al. 2009). We do not detect any galaxy with mass larger than $10^{11.6}
M_{\odot}$, and most of our  objects have stellar masses
between $10^{10.5}$ and $10^{11.3} M_{\odot}$. According to
 the S-COSMOS mass functions \citep{ilbert:09}, within this $M_{*}$ range, only less
than $\sim$ 40, 35 and 20 \% of all galaxies are ``quiescent''
(i.e. lie on the red sequence) for redshift ranges [1, 1.2], [1.2, 1.5]
and [1.5, 2.0], respectively. 

A similar conclusion about the average star-formation properties of
the type--1 AGN hosts can be drawn by studying the mass to light ratio
of the best fitting models. Figure~\ref{fig:mlk} shows the relation
between rest frame K band luminosity and total stellar mass. Once
again, we can compare these estimates of the $M/L_{K}$ ratio with
those measured by Ilbert et al. (2009) in the COSMOS IR-selected galaxy
sample (see Ilbert et al. 2009, appendix D). Taking into account the
$\approx$0.24 dex shift due to the different IMF choices, our mean Log$(M/L_{K})\sim -0.55$
follows more closely the expectations for the so-called blue-cloud
(star-forming) galaxies, but the nominal uncertainties remain large\footnote{We note here that the level of
  uncertainty of the SFR for our AGN hosts is such that we cannot
  clearly discriminate between what, in the galaxy formation jargon,
  is usually called ``blue cloud'' and ``green valley'' (see
  e.g. Silverman et al. 2008)}.

As a further test of our method, we have compared the total stellar
mass estimates with those derived by Jahnke et al. (2009) for a small
sample (18 objects) of X-ray selected AGN in the COSMOS field for
which simultaneous {\it HST}/ACS and {\it HST}/NICMOS observations
allow to derive mass to light ratios (and stellar masses) of the
resolved hosts. We find that the two independent methods are in broad
agreement with each other. 5/18 of them have also zCOSMOS spectra, and
are part of the sample described here (see the empty stars in
Fig.~\ref{fig:mbh_mk_mass}). 
Their total stellar masses calculated with the two methods
agree to better than 0.2 dex, with our estimates being consistently on
the lower side. We discuss this issue further in section~\ref{sec:our_results}.

\section{Virial black hole mass measurements} 
\label{sec:virial}
Due to the uncertainties in the actual geometry of the broad
line region of type 1 AGN (see e.g. Mc Lure and Dunlop 2001), and/or to
the different choice of absolute calibration of the black hole masses
of local AGN studied with reverberation mapping, 
there exists a large number of different formulae in the literature
that relate black hole mass, \MgII\ line width and continuum luminosity
(see e.g. Mc Gill et al. 2008, and references therein).
They can all be expressed in the form:
\begin{equation}
\label{eq:virial}
\log\frac{M_{BH}}{M_{\odot}}=A+\log(FWHM^2_{1000}(\lambda L_{3000,44})^{\beta})
\end{equation}
where $FWHM_{1000}$ is the FWHM of the line in units of 1000 \kmps,
and $\lambda L_{3000,44}$ is 
the continuum luminosity at 3000\AA\ in units of 10$^{44}$ erg $s^{-1}$.
For \MgII\ lines, Mc Lure and Dunlop (2004) proposed A=6.51 and $\beta=0.62$, based on a
fit to the radius-luminosity relation for AGN with reverberation
mapping and
total luminosity $\lambda L_{\lambda}>10^{44}$ erg/s, while Vestergaard
et al. (2009; in preparation; see also Trump et al. 2009) propose a 
different scaling (A=6.86, $\beta=0.5$). Each of these relations carries a significant scatter of about 0.3 dex
(McGill et al. 2008). The exponent $\beta$ is related to the empirical
calibration of the radius-luminosity relation \citep{kaspi:00}. A
recent re-analysis of reverberation mapping observations, 
fully accounting for host-galaxy light contamination \citep{bentz:09},
points towards a value of $0.45\la \beta \la 0.59$, consistent
with the simple expectations from photoionization models of the broad
line region ($\beta=0.5$).
Here we will adopt the 
relation derived by McGill et al. (2008), with A=6.77 and $\beta=0.47$. This was derived by
cross-calibrating a number of different estimators applied to BLAGN in
a redshift range where more than one broad emission line can be
observed simultaneously in optical spectra.
Additional systematic errors in the derived black hole masses introduced
by the use of any of these relations will be further discussed in
section~\ref{sec:bhmass} (see also the discussion in Treu et
al. 2007).

We do not take into account possible effects due to radiation
pressure on the Broad Line Region (BLR), that could lead to systematic
under-estimate of the black hole mass for the objects with the higher
Eddington ratio. Such an effect has been estimated
empirically by Marconi et al. (2008) for the H$\beta$ broad emission
in a local AGN sample. No calibration is currently available for
\MgII\ emission lines. Moreover, the size of the possible radiation
pressure correction is  expected to scale linearly with the inverse of
the Eddington ratio, and should not be too large for the
moderate-luminosity AGN in
our sample (see section~\ref{sec:bhmass} below). We thus made the
conservative choice not to include any radiation pressure correction
to the virial relation (\ref{eq:virial}). As we will discuss at length
in the following, this corresponds to {\it minimize} the amount of
possible evolution detected in the average $M_{\rm BH}/M_{*}$ ratio as
a function of redshift.

In the following subsections, we will describe the method used to
measure both line width and continuum luminosity needed to apply
eq.~(\ref{eq:virial}).
The measured $FWHM_{1000}$, $L_{3000}$, together with their 1$\sigma$ errors, are
given in Table~\ref{tab:agn}.

\subsection{FWHM measurement}
Type--1 (un-absorbed) AGN spectra in the wavelength region of interest
are usually characterized by a power-law continuum, of the form
$f_{\lambda} \propto \lambda^{\alpha}$, broad line emission from \MgII\ plus a
complex of \FeII\ emission lines, also broadened at the typical
velocities of the broad line region (Boroson and Green 1992;
Vestergaard \& Wilkes 2001). Accurate subtraction of the broad
\FeII\ features is thus an important step in the process of obtaining
broad \MgII\ line widths. It is not always straightforward to keep
track of the \FeII\ subtraction technique in previous
studies high-redshift scaling relations. For example, Peng et
al. (2006a) erroneously report that Mc Lure and
Jarvis (2002) did not performed any \FeII\ subtraction, contrary to
what stated in section 3.2 of Mc Lure and Jarvis (2002). For the sake
of completeness, we briefly discuss in section~\ref{sec:bias_bhmass}
what would be the typical systematic effect of not removing an Iron
template from the spectra.

We adopt here the theoretical \FeII\ template calculated by
Bruhweiler \& Verner (2008), calibrated by fitting the Seyfert
1 galaxy IZw1 spectrum (model Fe\_d11-m20-20.5 available at {\tt http://iacs.cua.edu/people/verner/FeII}). 
In order to apply the iron template to the spectra, the line width of the template must 
be matched to that of the AGN spectrum. We achieve this by means of an
iterative procedure, as described below.

First of all, we convolved the original \FeII\ template 
with gaussian functions of different widths, ranging from $v=1000$\kmps
to $v=15000$\kmps\ in step of 250\kmps to produce a grid of broadened templates.  
From the original AGN spectrum we then derived: (a) a first rough estimation of 
the FWHM of the \MgII\ line using a single gaussian fit with \textit{iraf-splot} 
package and (b) the mean values of the continuum flux and the corresponding errors 
(derived from the noise spectra) in the following wavelength 
windows:  
[2660\AA-2700\AA]; [2930\AA-2970\AA]; [2715\AA-2750\AA]; [2850\AA-2885\AA] 
and [2980\AA-3020\AA]\footnote{For a handful of objects (5) with relatively small
FWHM, a significant improvement in the fit is obtained by shifting
the two central windows for continuum plus FeII fitting to
[2730\AA-2770\AA]; [2830\AA-2870\AA], closer to the most prominent
emission peaks of the FeII complex.}.
 We assumed that in 
these wavelength ranges, the spectrum can be completely described by 
a combination of power law and
\FeII\ emission smeared by a velocity width $v$ equal to the one derived from the 
\MgII\ emission line 
\begin{equation}
F(\lambda)=a\lambda^{\alpha}+b \times \FeII(\lambda,v).
\end{equation}
In order to find the best fitting model, we then performed a chi-square minimization 
and we thus derive the three parameters $\alpha$, $a$ and $b$.
Finally, we subtracted the best fit power-law plus \FeII\ model 
to the original spectrum obtaining a new spectrum 
that contains only the \MgII\ emission line. This is done using the 
\textit{iraf-sarith} package.

We then measured the FWHM of the line, modeled using one, two or three gaussians 
(absorption and/or emission) and we choose the best solution according to best 
reduced $\chi^2$ computed in the [2650\AA-2950\AA] range. 
The best fit model for the emission line spectrum is thus given by:
$M_{\lambda}=\sum_{i=0}^{N} a_i G_{\lambda}(\lambda_{peak,i}, \sigma_i)$,
where N is the number of gaussian components chosen by the fit. For each gaussian 
component $G_{\lambda}$, $a_i$ is the intensity (positive for emission, negative for absorption), 
$\lambda_{peak,i}$ the wavelength of the peak and $\sigma_i$ the width. 
Each of these quantities carries a statistical error provided by the \textit{splot-iraf} package.

\begin{figure}[ht]
\includegraphics[scale=.27,angle=0]{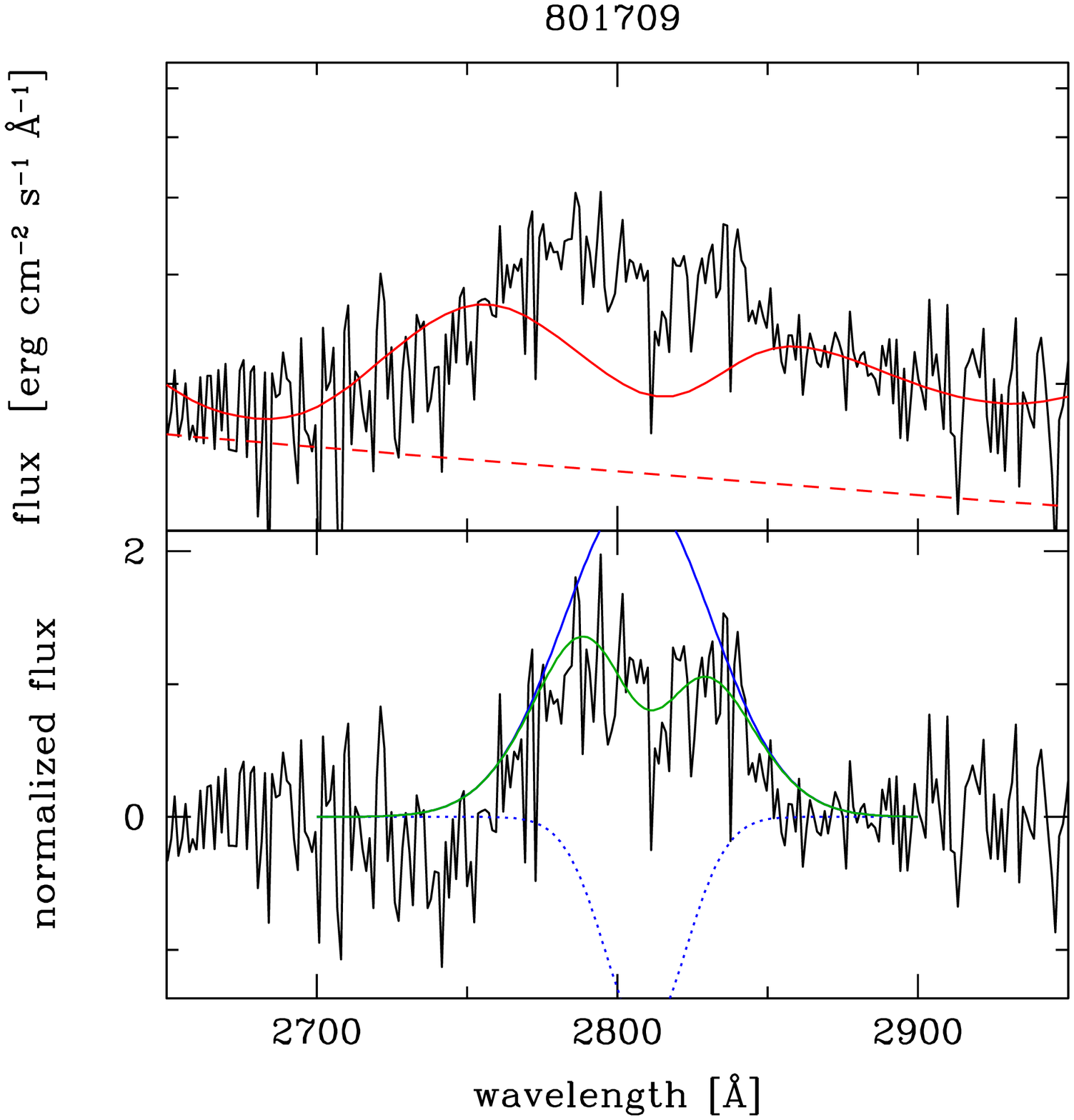}
\includegraphics[scale=.27,angle=0]{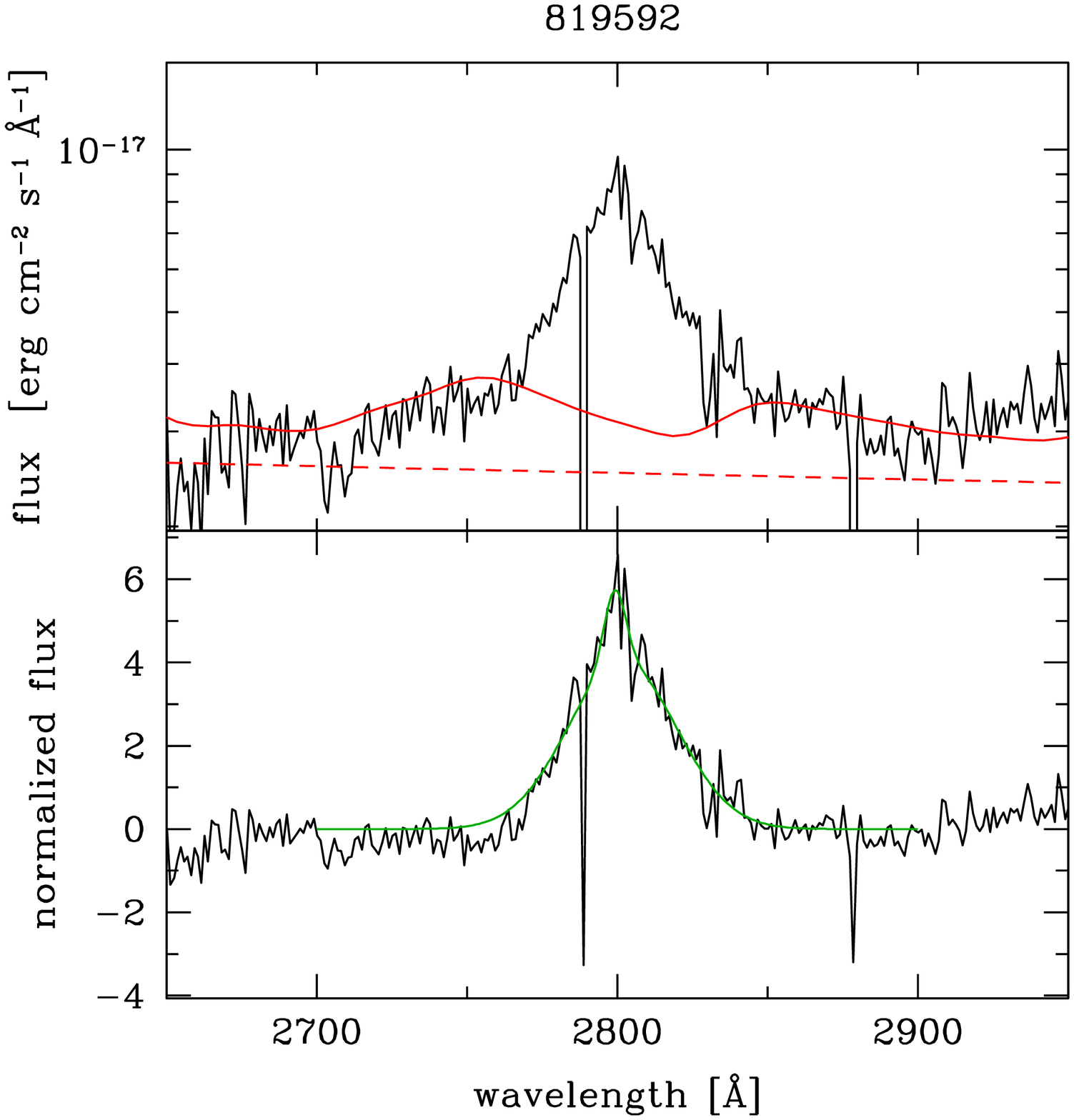}
\includegraphics[scale=.27,angle=0]{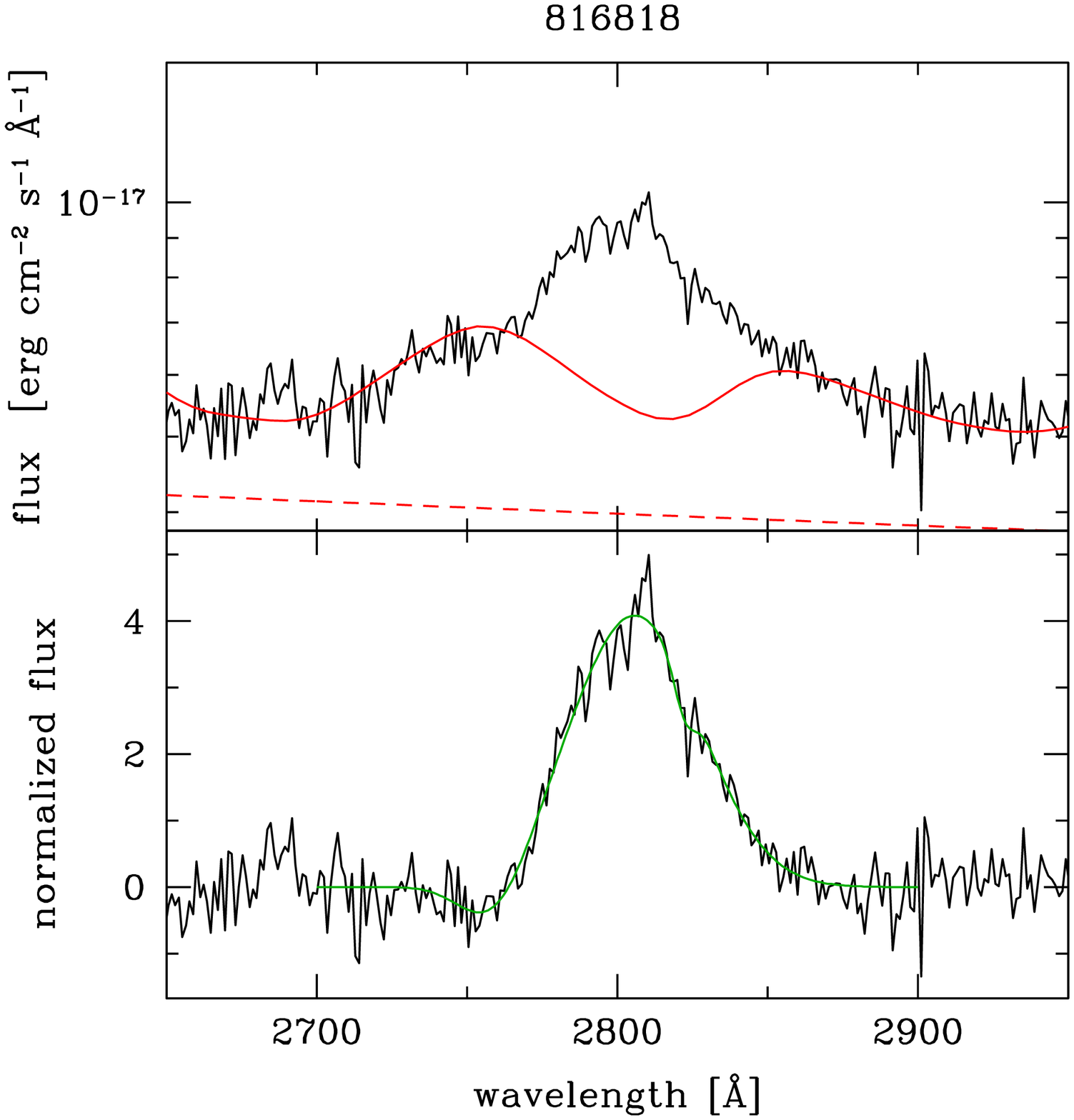}\\
\includegraphics[scale=.27,angle=0]{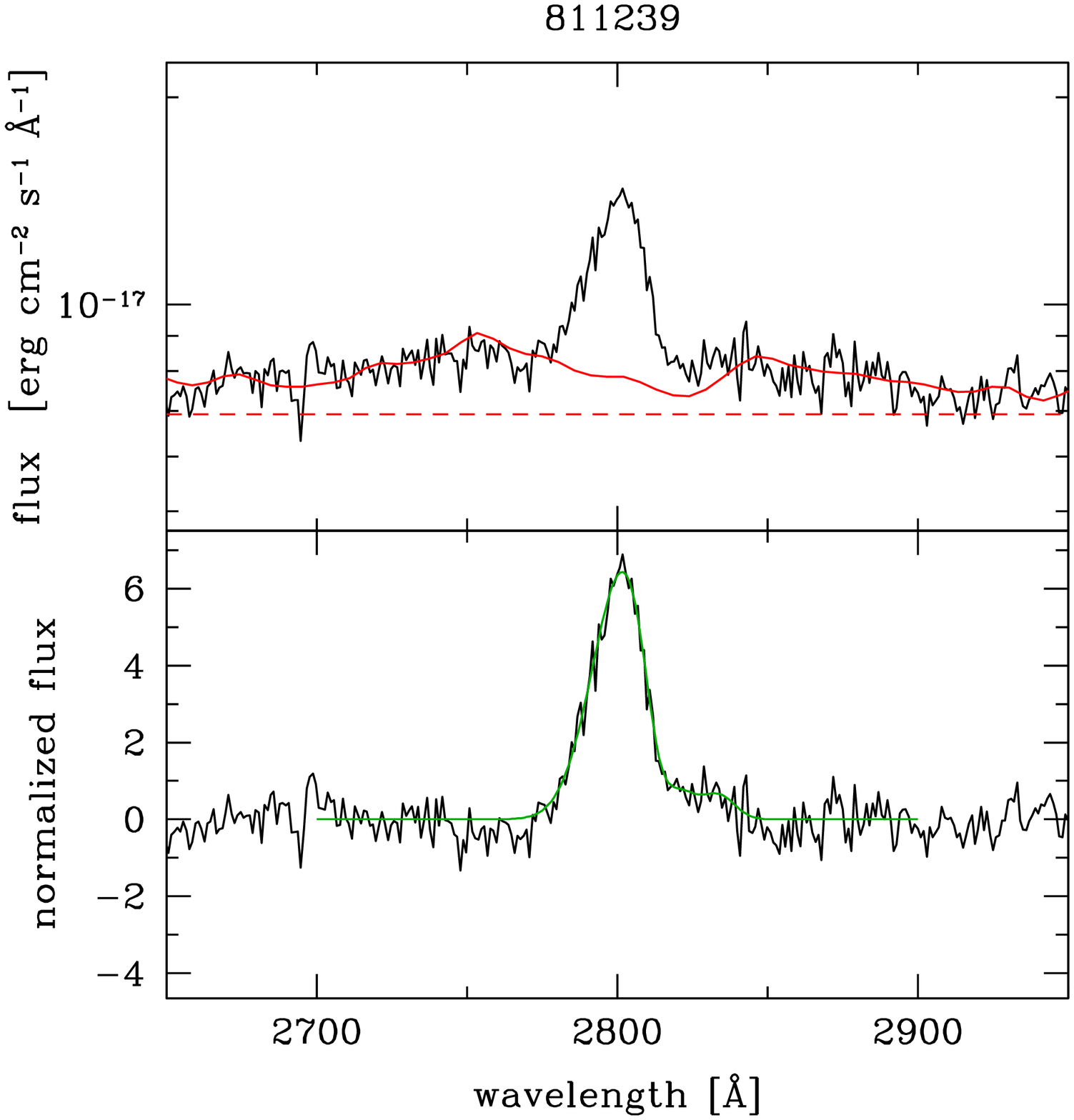}
\includegraphics[scale=.27,angle=0]{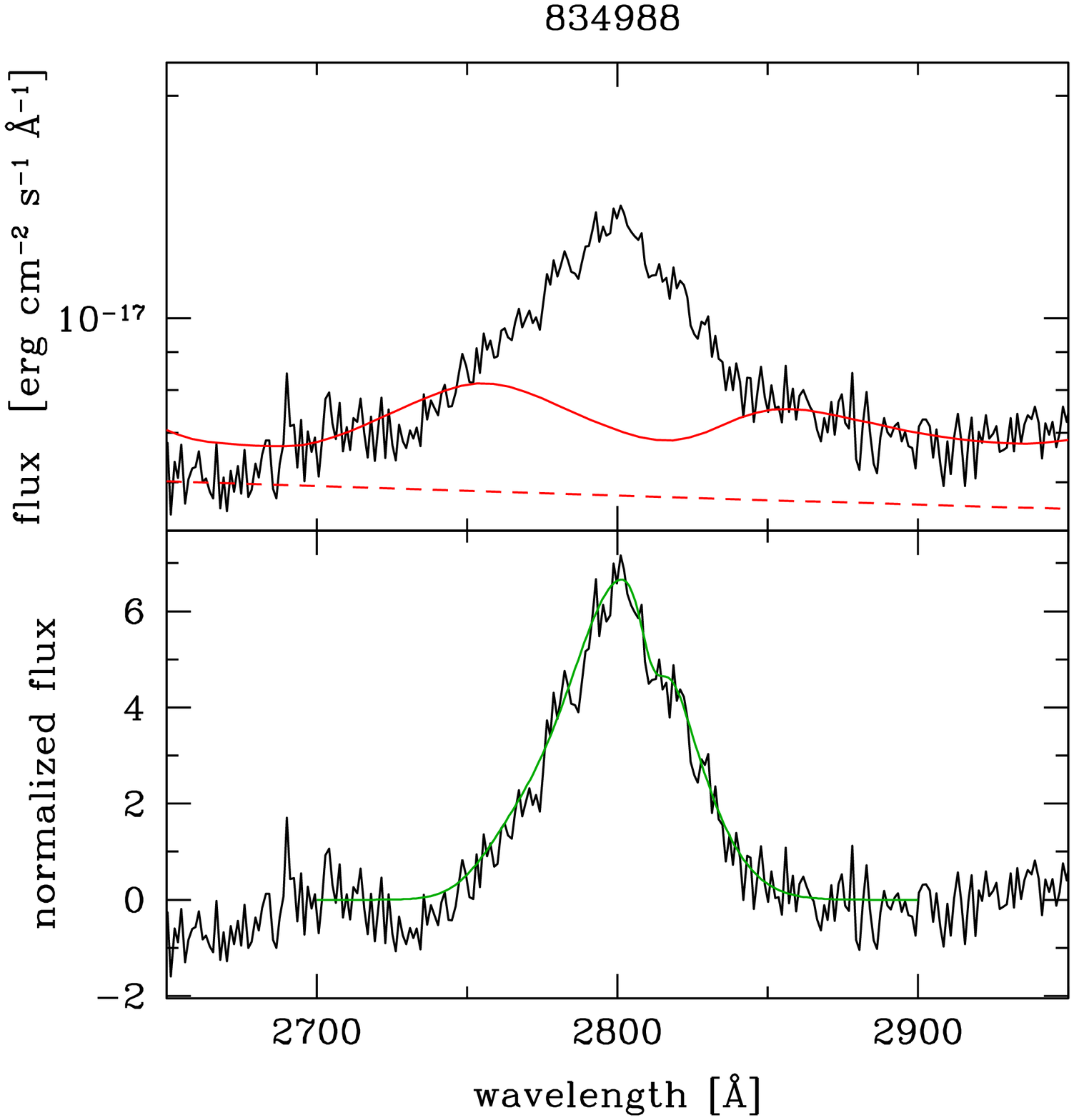}
\includegraphics[scale=.27,angle=0]{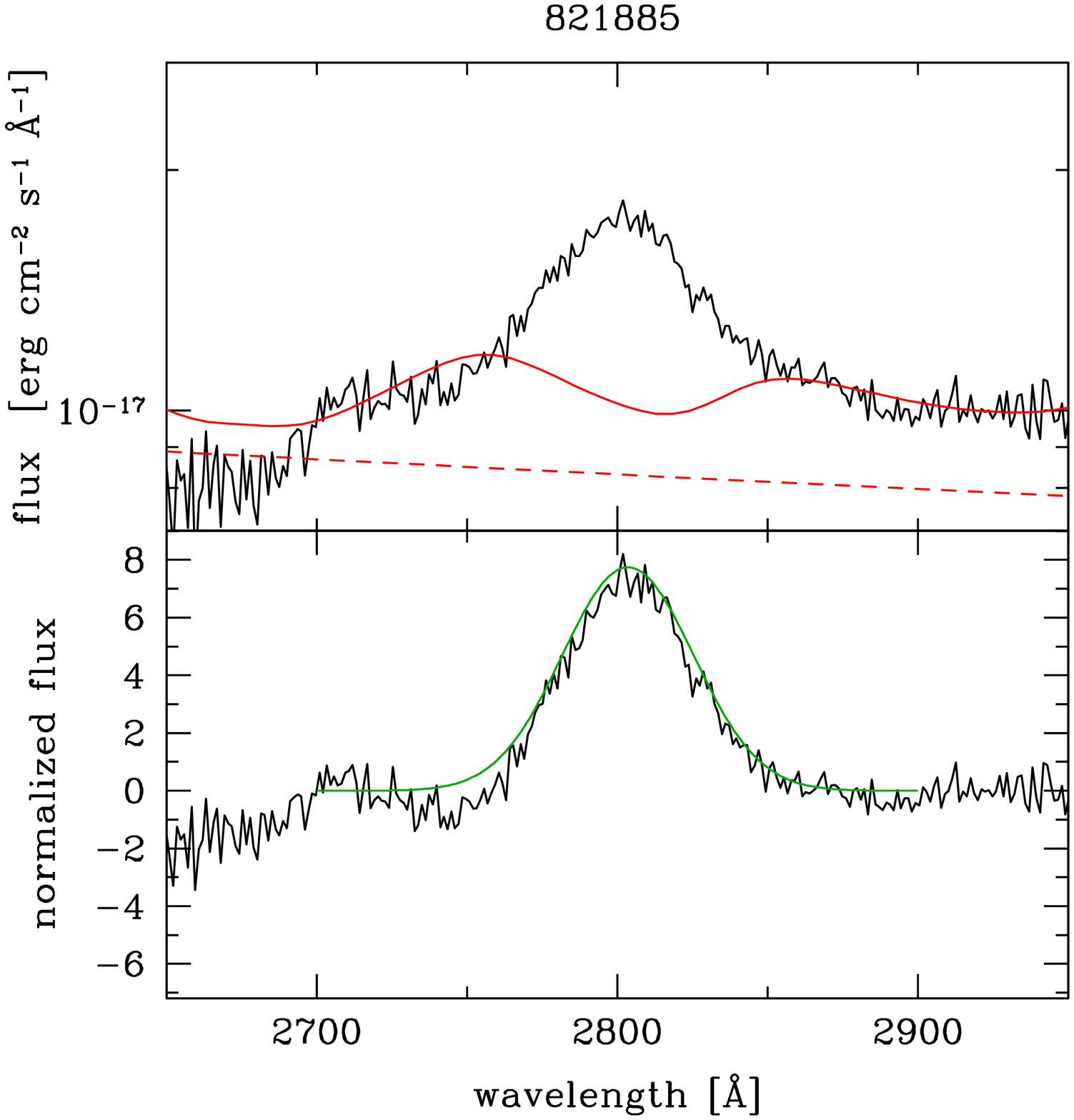}\\
\includegraphics[scale=.27,angle=0]{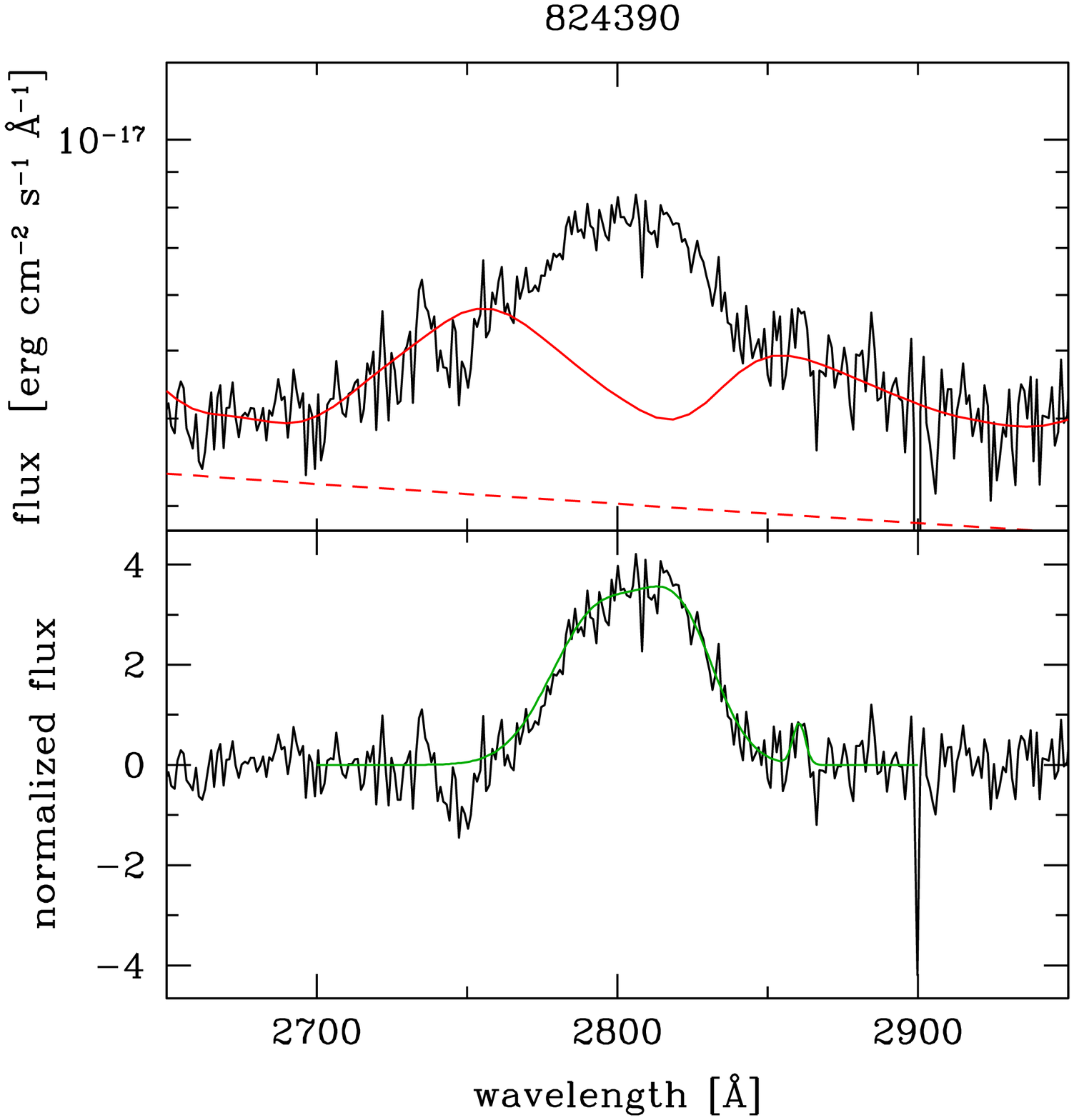}
\includegraphics[scale=.27,angle=0]{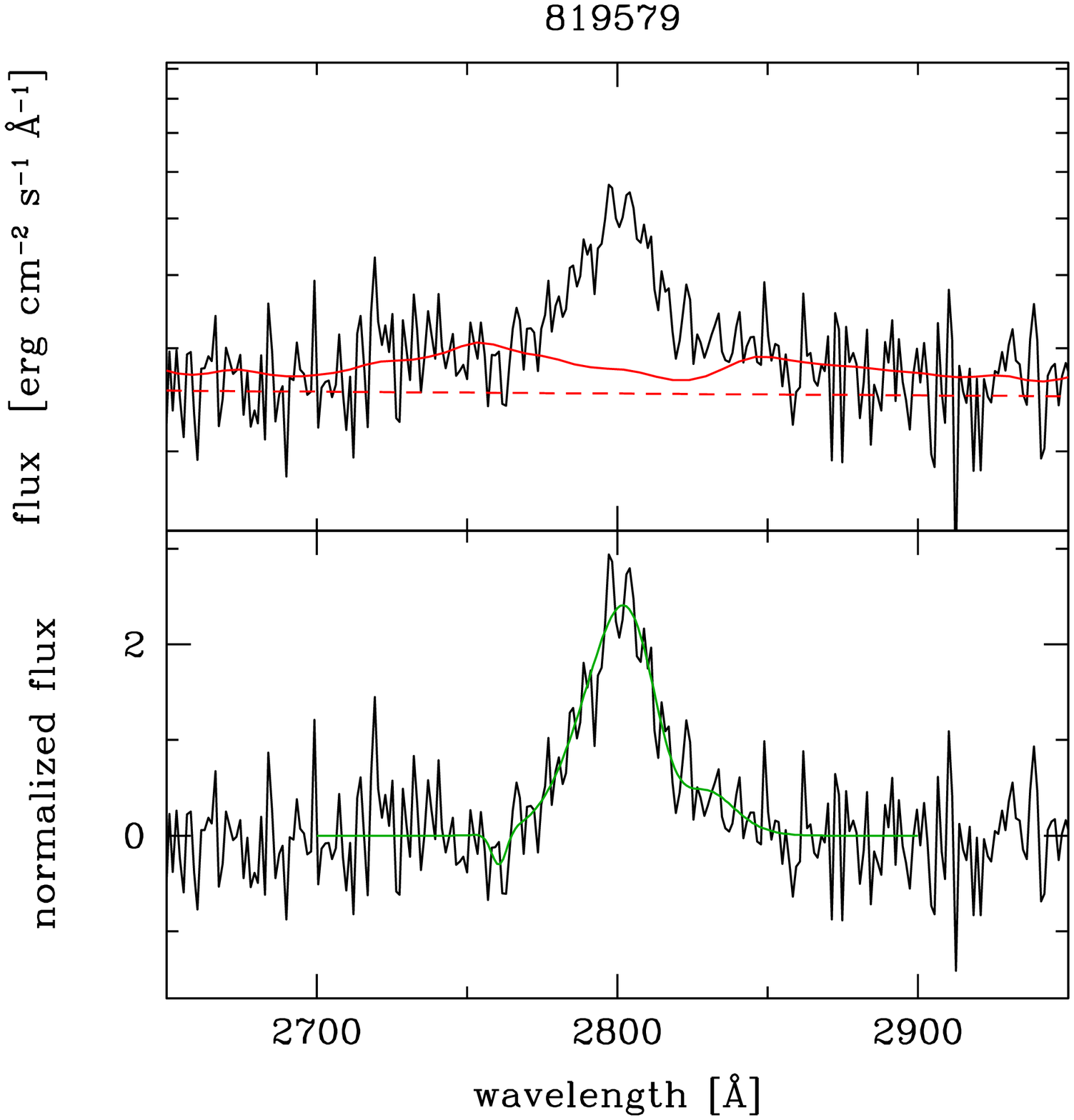}
\includegraphics[scale=.27,angle=0]{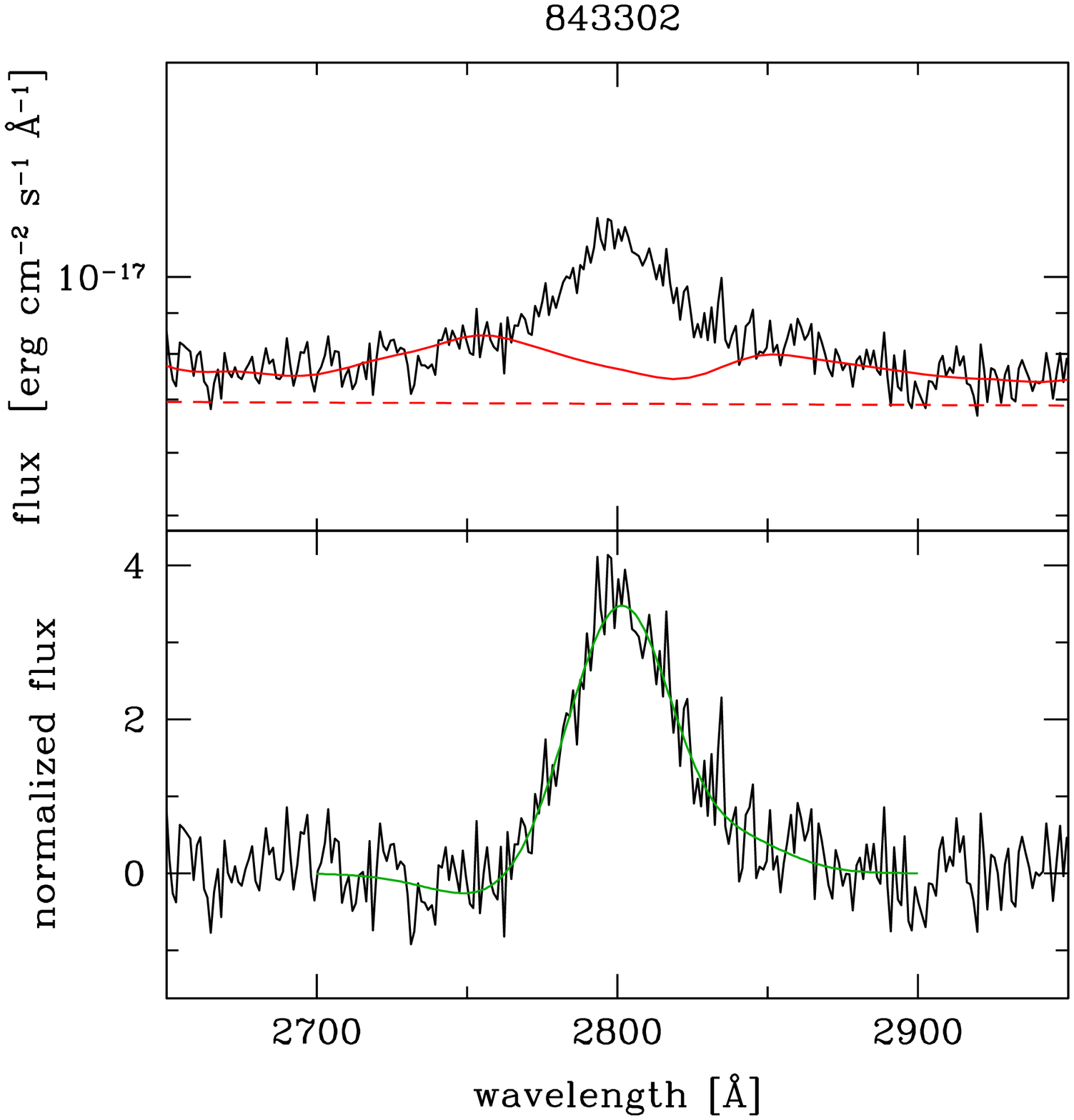}
\caption{A few examples of the spectral fitting of the \MgII\ spectral
  region. For each object, the \textit{upper panel} shows  the fit to
  the power-law continuum (red dotted line) and the \FeII\ emission
  (red solid line) that we then subtract to the spectrum. 
In the \textit{bottom panels} we show instead the same spectrum after
the \FeII\ and continuum subtraction, with the solid green line is the
final best fit. Note that some of the apparent absorption components
redwards of the line are not
physical, and are required by the fitting routing to compensate for an
incorrect representation of the \FeII\ emission. In the top left
panel, 801709 is one of the three objects in the sample showing signs
of absorption in the \MgII\ line region. They are highlighted as blue
open circles in Fig.~\ref{fig:mbh_mk_mass}; for them, the FWHM is
computed from the emission components only.} 
\label{fig:bl_fits}
\end{figure}

We then compute the FWHM of the model emission line complex, and the 
error on the total FWHM, $\sigma_{fwhm}$ is computed by propagating the errors on the single 
gaussian components of the fit.
Using the new FWHM determination for the line, we then iterated the
whole procedure (we subtract the new broadened iron template chosen according to the 
FWHM measured from the original spectrum), until the fits converge and the final 
measure of the FWHM is stable. 
The final FWHM measurement is corrected for the finite spectrograph 
resolution assuming that $FWHM^2_{intrinsic}=FWHM^2_{oss}-\frac{\lambda_{eff}^2}{R^2}$
where R is the mean instrumental resolution that for the zCOSMOS
spectra is $\simeq$580.
The best fit decomposition of the \MgII\ region of the spectra is
shown in Figure~\ref{fig:bl_fits} for the same nine AGN whose SED
decomposition is shown in Fig.~\ref{fig:sed_decompose}. For three objects
the fitting routine requires the presence of absorption in the \MgII\
line region (one is shown in the top left panel in figure~\ref{fig:bl_fits}). They are
marked as blue open circles in figure~\ref{fig:mbh_mk_mass}. Some residual
errors due to over-subtraction of the iron template red-wards of the
\MgII\ emission are apparent in some cases. It is clear that a single
FeII template may not be adequate to fully describe this complex
emission in all objects. A detailed study of the properties of the
FeII emission requires higher signal-to-noise spectra, and is beyond
the scope of the present paper.

\subsection{3000\AA\  and bolometric Luminosity}
From the best fitting power-law continuum we derive an estimate of the
AGN luminosity at 3000\AA. 
Our points have been corrected for aperture effects normalizing the
i-band fluxes measured in the VIMOS spectra with the observed ACS
i-band photometry.
Monochromatic continuum  luminosities at 3000\AA\ were then calculated from the average best fit 
continuum flux rescaled in the 2980-3020 \AA\ rest frame. 
The error on the continuum luminosity is obtained from the average of
the noise spectrum in 
the same wavelength range.

\begin{figure}[ht]
\epsscale{0.75}
\plotone{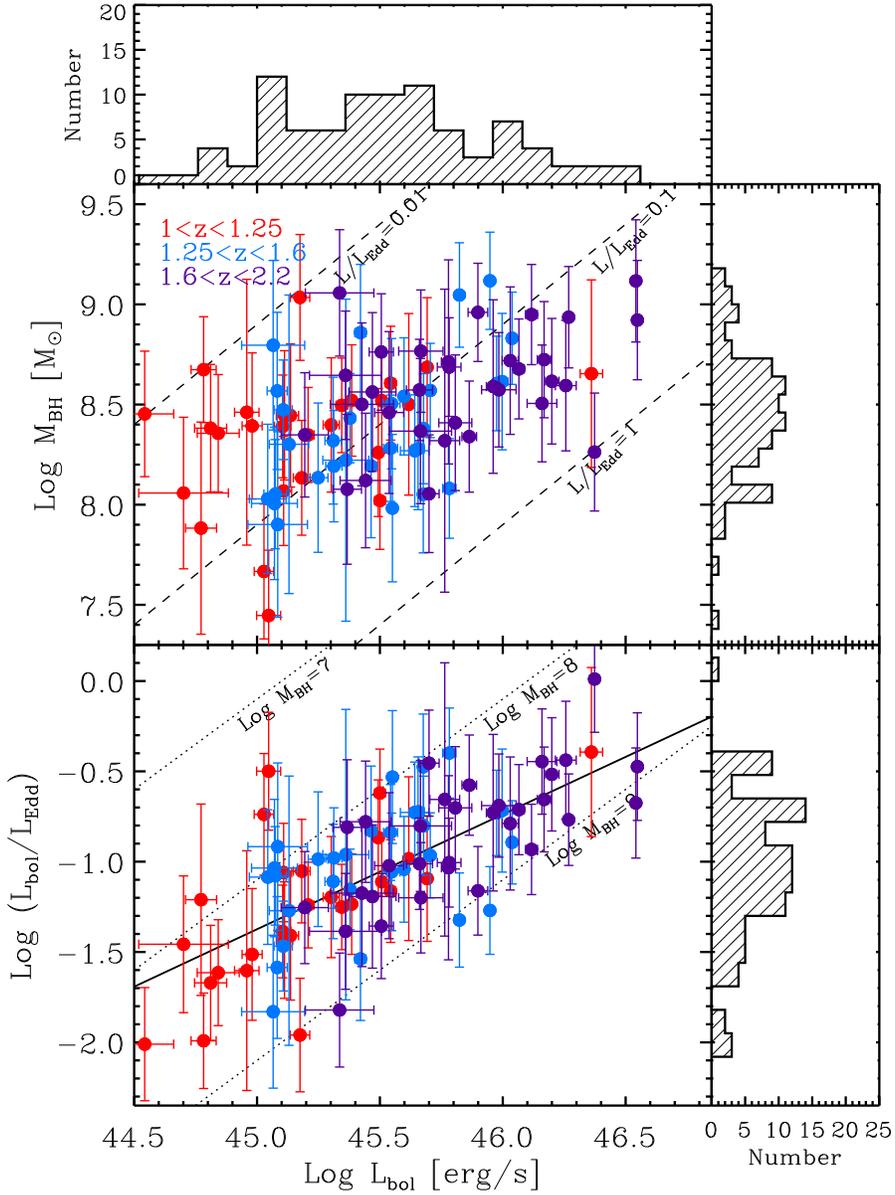}
\caption{{\it Large Top Panel}: Black hole mass vs. bolometric luminosity
  for the zCOSMOS BLAGN in the redshift range $1<z<2.2$. Different
  colors correspond to different redshift ranges, while dashed lines
  mark the loci of constant Eddington ratios. {\it Large Bottom Panel}:
  Eddington ratio vs. bolometric 
luminosity for the same objects. Dotted lines mark the loci of
constant BH mass, while the black solid line is the best fit linear
regression to the data points, given by ${\rm Log}(L_{\rm bol}/L_{\rm
  Edd})=-1.38+0.64{\rm Log}L_{\rm bol,45}$, where $L_{\rm bol,45}$ is
the bolometric luminosity in units of 10$^{45}$ erg s$^{-1}$. The
small panels in the top and right hand side display the distributions
of bolometric luminosity, black hole mass and Eddington ratio, respectively.} 
\label{fig:edd_l}
\end{figure}

From the measured total $\lambda L_{\rm 3000}$ AGN luminosity we
derive the bolometric one using the luminosity-dependent bolometric correction factor
$f_{\rm bol}$ of Hopkins et al. (2007). The distribution of bolometric
luminosities for the type--1 AGN in our sample is shown in the upper
panel of Figure~\ref{fig:edd_l}. Assuming a standard radiative
efficiency of $\epsilon_{\rm rad}=0.1$, the median accretion rate onto the black
holes, $\dot M=L_{\rm bol}/\epsilon_{\rm rad}c^2$, is of the order of
0.4 $M_{\odot}/$yr.


\subsection{Black hole masses and Eddington ratio distribution}
\label{sec:bhmass}
We have tested the impact of the choice of single-epoch
virial formula for the black hole mass uncertainty. The typical spread
in Log$M_{\rm BH}$ among the three different estimators discussed in section~\ref{sec:virial}
is of the order of 0.2 dex, similar to the observed scatter
in the virial relations themselves (Vestergaard \& Peterson 2006;
McGill et al. 2008). In the following, we fix the systematic
uncertainty in the Log$M_{\rm BH}$ determination to 0.2 dex, and for
our statistical analysis of the scaling relation evolution, we assign
to each black hole mass measurement an error given by the sum of
the statistical and systematic uncertainties.

Figure~\ref{fig:edd_l} shows the estimated black hole masses vs. the
bolometric luminosities for all the objects in our sample (upper
panel) as well as their location in the Luminosity-Eddington ratio ($\lambda\equiv
L_{\rm bol}/L_{\rm Edd}$, where $L_{\rm Edd}=1.3\times 10^{38} M_{\rm
  BH}/M_{\odot}$ is the Eddington luminosity) plane in the lower panel. 
As typical for optically selected samples of type--1 AGN/QSOs
\citep[and references therein]{kollmeier:06,gavignaud:08,trump:09}, 
the distribution of Eddington ratios is quite narrowly distributed,
with a median of $\approx 0.1$.

Analogously to what found in BLAGN samples selected in similar redshift
ranges and at comparable depths (VVDS, Gavignaud et al. 2008), we also observe
a trend of increasing Eddington ratio with increasing bolometric
luminosity of the AGN (see also Netzer et al. 2007, for a sample of
higher-redshift QSOs). Fitting a straight line we find ${\rm
  Log}\lambda \propto 0.64\, {\rm Log}L_{\rm bol}$, but the slope of
such a relation depends critically on the exponent $\beta$ in the adopted
virial relation~(\ref{eq:virial}). As already pointed
out in Gavignaud et al. (2008), if we chose the McLure and Dunlop
(2004) formula (with $\beta$=0.62), we would obtain a shallower slope:  ${\rm
  Log}\lambda \propto 0.47\, {\rm Log}L_{\rm bol}$.
Selection effects could certainly be playing a role in
determining the distribution of sources in the $\lambda-L_{\rm bol}$
plane (for example, objects in the lower right corner, i.e. massive
black holes at low accretion rates, probe the massive end of the SMBH
mass function, which is rapidly declining in this redshift range, see
e.g. Merloni \& Heinz 2008). The detailed distribution of AGN
lifetimes as a function of luminosity can also be responsible for
the observed trends, as suggested by some numerical models for QSO
lightcurves (Hopkins and Hernquist 2009). For further discussion of
possible selection and/or systematics effects, we refer the reader to the
work of Trump et al. (2009), who have studied in greater detail
the larger sample of XMM-COSMOS AGN with IMACS spectroscopy.

\section{Scaling relations and their evolution}
\label{sec:scaling_evol} 
In this section, we quantify the amount of evolution (if
any) in the scaling relations between nuclear black holes and host
galaxy properties observed locally.

\begin{figure}[ht]
\epsscale{0.8}
\plotone{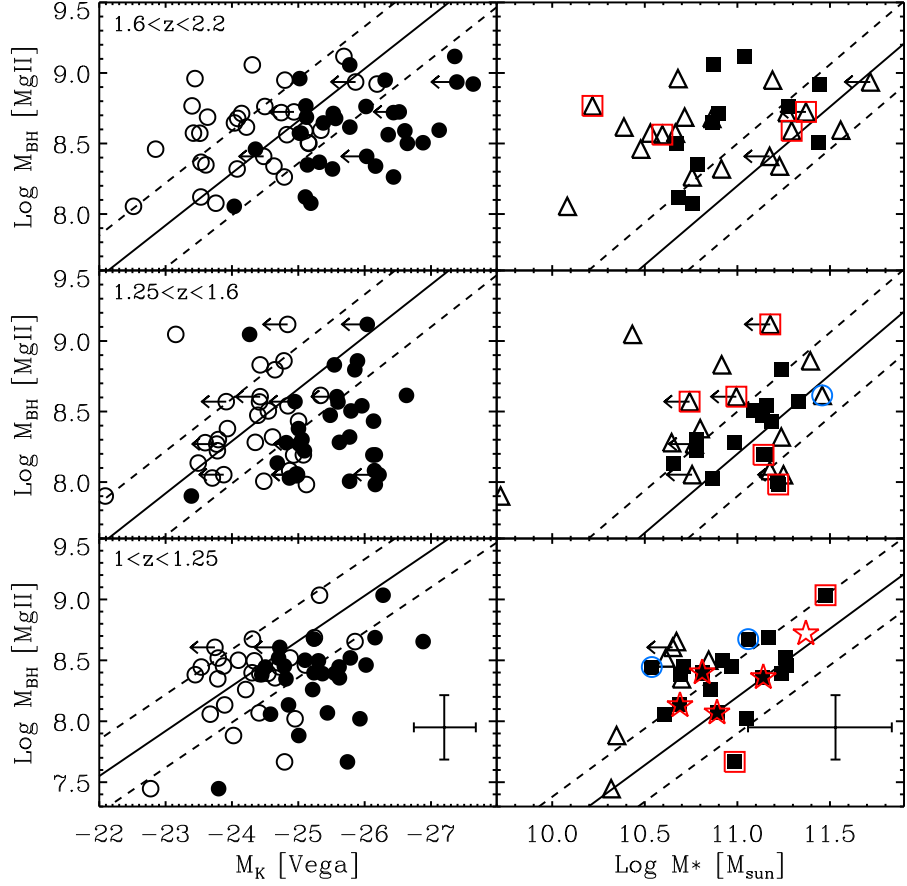}
\caption{Scaling relations  for
  zCOSMOS type--1 AGN in the redshift range $1<z<2.2$. Each row show a
  different redshift interval (lowest $1< z < 1.25$; middle $1.25 < z
  < 1.6$; top $1.6 < z < 2.2$). {\it Left panels}: Black hole mass host K-band absolute magnitude relation.  Filled
symbols represent measurements, leftwards arrows upper limits on
the host luminosity. The black solid line is the best fit to the
Graham (2007) local spheroids sample relation, with dashed lines
marking a $\pm$0.3 dex offset. The typical error bars of our measurements
are shown as black cross in the lower right corner. Black open circles
mark the location of our galaxies when passively evolved down to $z=0$
assuming a formation redshift of $z_f=3$. 
{\it Right panels}: Black hole mass host stellar mass relation.
Open triangles (filled squares) denote the objects
with low ($<0.33$) and high ($>0.33$)
Galaxy to AGN luminosity ratio in the rest frame K-band, respectively.
Symbols with leftwards arrows represent upper limits on
the host mass. The black solid line is the best fit to the
\cite{haering:04} local spheroids sample relation, with dashed lines
marking a $\pm$0.3 dex offset. 
Red squares mark the objects detected in
the radio by VLA at 1.4 GHz, while red stars mark the location of the 5
zCOSMOS AGN in the Jahnke et al. (2009) sample. Blue circles mark the
objects with absorption features in the \MgII\ line.
The typical error bars of our measurements
are shown as black cross in the lower right corner.} 
\label{fig:mbh_mk_mass}
\end{figure}

We begin by showing in the left panel of Figure~\ref{fig:mbh_mk_mass}
the location of our AGN (filled circles) in the Log $
M_{\rm BH}$-$M_K$ plane. As a reference, we show there the best
fit relation derived from local inactive galaxies by Graham (2007) and
given by 
\begin{equation}
\label{eq:graham}
{\rm Log} M_{\rm BH}=8.29-0.37(M_K+24)
\end{equation} 
(to guide the eye, we show also
the same relation offset by $\pm$0.33 dex, the total scatter
in the Graham (2007) relation). Our objects are shifted towards
brighter hosts and/or smaller black hole masses with respect to the
$z=0$ relation.
The objects do not seem to obey any tight relation between black hole
mass and $M_K$, although the range of $M_{\rm BH}$ probed is
relatively limited. If one tried, for the sake of comparison, to fit
the sample with a relation with the same slope as in
eq.(\ref{eq:graham}), the best fit normalization would be shifted by
$\approx 0.28$ dex, and the intrinsic scatter would be as large as
0.44 dex. 
We have also measured the offset,
$\Gamma_K(z)$ (here defined as the
projected distance in the Log$M_{\rm BH}$-Log$L_{\rm K}$ plane), of
each observed point from the
local relation and studied its evolution as a function of redshift. 
The best fit obtained imposing the functional form
$\Gamma_K(z)=\delta_1 {\rm Log}(1+z)$, gives $\delta_1=-0.73 \pm
0.08$ (these results are unchanged if we instead adopt the $M_K$ derived by fitting the SED with
the P07 template).
The fit are performed taking into account both errors and lower limits
on $\Gamma$, using a Monte Carlo approach, 
described in Bianchi et al. (2007).

Such an evolution is somewhat stronger than that observed by Peng et al. (2006b),
where the R-band luminosity of 30 lensed and 20 non-lensed QSO hosts
was measured based on detailed HST image modeling of $1<z<4.5$
quasars. Their QSOs apparently lie almost exactly on the same observed $M_{\rm
  BH}$-$L_{\rm R}$ relation as their $z=0$ relic counterparts.
However, as already noted by Peng et al. (2006b), high
redshift galaxies shall have a different mass to light ratio as compared
to their $z=0$ descendants, at the very least because of passive
evolution of the stellar population,
in the extreme case of non-star forming galaxies. 
This can be simply accounted for by passively evolving the rest-frame
K-band luminosities down to $z=0$ to allow a more direct comparison
with the local relation. We have done this using once again the
Bruzual and Charlot (2003) template libraries, and assuming all
objects formed all their stars in burst at $z_f=3$. Their local
descendants (passively evolved) would have dimmed to the magnitudes
indicated by the black empty circles in the right panels of
Fig.~\ref{fig:mbh_mk_mass}, which are now broadly consistent with the
local $M_{\rm BH}-M_K$ relation, but slightly offset towards large
black hole to host galaxy ratios. 
Such a shift towards a positive offset from the local relation would in
fact be even stronger if indeed the host galaxies of our AGN sample were dominated, at
least statistically, by actively star forming galaxies, as
discussed earlier in section~\ref{sec:mstar}.

This is indeed what happens when we consider the relationship between measured black hole masses
and host total stellar masses, $M_{*}$, obtained from the BC03 fits. 
In the right panels of Figure~\ref{fig:mbh_mk_mass} we show the
location of 89 zCOSMOS AGN
in the Log$M_{\rm BH}$-Log$M_{*}$ plane.  In these panels, we have
separated the objects on the basis of the measured $f_{\rm gal,K}$
host-to-AGN luminosity ratio in the rest frame K-band (with open triangles
having $f_{\rm gal,K}<0.34$ and filled squares $f_{\rm gal,K}>0.34$). 
 As a reference, a solid line
shows the local best fit relation between black holes and spheroids as
derived by \cite{haering:04}:  
\begin{equation}
\label{eq:haering04}
{\rm Log} M_{\rm BH}=-4.12+1.12({\rm Log}M_{*})
\end{equation}

Our objects now show a modest, but clear, offset from the local
relation, especially at the high black hole mass end (see also
Figure~\ref{fig:dmbh_mbh}) and in the highest redshift bin (upper
right panel of Fig.~\ref{fig:mbh_mk_mass}).
Radio-detected AGN (squares) appear to be distributed similarly to the
rest of the population, but low number statistics prevent us from
reaching any firmer conclusion on their properties.

As our method is affected by substantial uncertainties in both black
hole and host galaxy mass, and none of the two can be treated as truly
independent variable, we chose to measure the deviation of our
data-set from the local scaling relation by measuring the distance
$\Delta {\rm Log}(M_{\rm BH}/M_{*})$ of
each point from the \cite{haering:04} relation, perpendicular to the
relation itself\footnote{Given the slope in the \cite{haering:04}
  relation $A=1.12$, the measured offset multiplied by
  $S\equiv\sqrt{1+A^2}\simeq 1.5$ gives the increase in black hole mass
  $\Delta M_{\rm BH}$ given a host galaxy mass, compared to the local
  value. This should be kept in mind when comparing with results from
  previous works, which usually measure the offset from the scaling
  relations in terms of ``excess black hole mass'', i.e. vertically in
Fig.~\ref{fig:mbh_mk_mass}\label{foot:6}}.
 
\begin{figure}[ht]
\epsscale{1.0}
\plotone{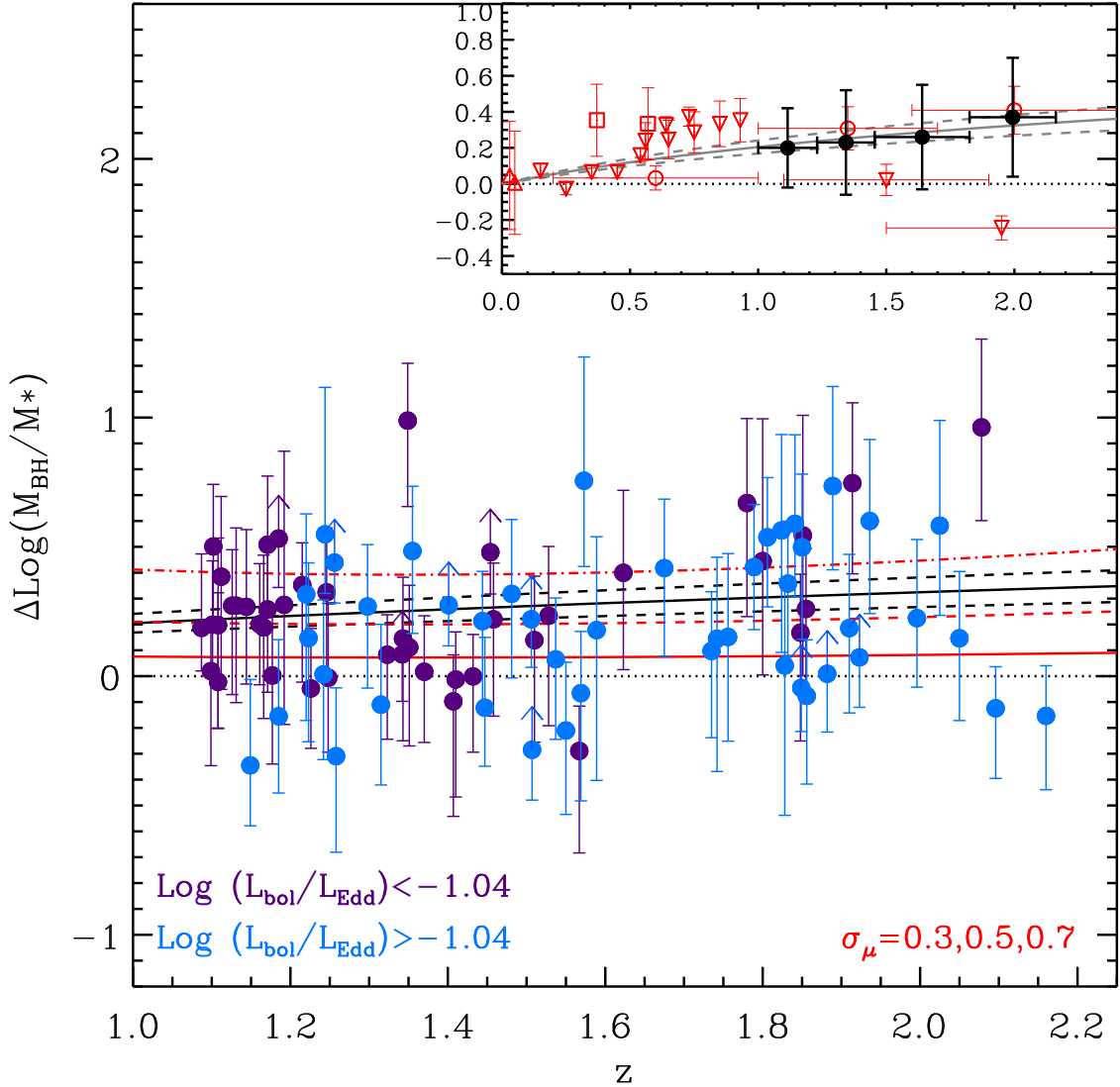}
\caption{Redshift evolution of the offset measured for our type--1 AGN from
  the local $M_{\rm BH}-M_{*}$ relation. Different colors identify different
ranges of Eddington ratios (purple circles Log $(L_{\rm bol}/L_{\rm
  Edd})<-1.$, and light blue ones Log $(L_{\rm bol}/L_{\rm
  Edd})>-1.$) with upwards arrows representing upper limits on 
the host mass. The offset is calculated as the
distance of each point to the \cite{haering:04} correlation. Solid black
line shows the best fit obtained assuming an evolution of the form
$\Delta {\rm Log}(M_{\rm BH}/M_{*})(z)=\delta_2{\rm Log}(1+z)$; for which we
found $\delta_2=0.68 \pm 0.12$. The red lines show the bias due to the intrinsic
scatter in the scaling relation to be expected even if they are universal. Solid line
is for an intrinsic scatter of 0.3 dex; dashed of 0.5 dex; dot-dashed
of 0.7 dex (see text for details). In the inset, we show a comparison
of our data (black circles)
with data from the literature, plotted as green open symbols: triangles are
from Salviander et al. (2007, low-z) and Shields et al. (2003,
high-z); squares from Woo et al. (2008) and circles from Peng et al. (2006b).} 
\label{fig:dmbh_z_mass}
\end{figure}

Figure~\ref{fig:dmbh_z_mass} shows the measured offset of each point from the
local relation as a function of redshift. The black solid line
shows the best fit obtained imposing the functional form
$\Delta {\rm Log}(M_{\rm BH}/M_{*})(z)=\delta_2 {\rm Log}(1+z)$, where we find
$\delta_2=0.68 \pm 0.12$. Also in this case we have used a Monte Carlo
simulation to derive the best fit linear regression coefficients;
lower limits were treated as if the true value of $\Delta {\rm Log}(M_{\rm BH}/M_{*})$ were
uniformly distributed up to a common value of 1.2.

The inset of Fig.~\ref{fig:dmbh_z_mass} shows
our dataset and best fit evolution, in black, as compared to a
number of estimates at lower or comparable redshift.
The best-fit evolution from the zCOSMOS data 
is in reasonable agreement with previous estimates both at lower
(Salviander et al. 2007) and at higher redshift (Peng et
al. 2006b). 
We note here that the significant amount of scatter in our dataset
translates into a relatively weak statistical significance of the
measured offset (see the
binned points in the inset of Fig.~\ref{fig:dmbh_z_mass}). Moreover,
any redshift dependence {\it within our sample only} is not statistically significant.
Although weaker, the evolution we measure is also marginally
consistent with that observed by Treu et al. (2007) and Woo et
al. (2008) from their sample of Seyfert galaxies at $z=0.36$
and $z=0.57$.

It is interesting to notice here that the majority of observational
data-points for AGN sample do indeed show a broadly consistent amount
of offset at all redshifts probed. This might suggest that intrinsic
differences in the SMBH/host galaxy relation between active and
inactive galaxies could play an important role besides any genuine
cosmological evolution.
Large, uniformly selected samples of AGN hosts, spanning a larger
redshift range than probed here (e.g. Decarli et al. 2009),
as well as accurate comparisons of scaling relations for active
(reverberation mapped) and
non-active galaxies at low-$z$ (see e.g. Onken et al. 2004), are and
will be very important in disentangling
true redshift evolution from other systematic differences with the
local samples.

Finally, no significant trend is found by dividing our sample into fast and
slow accretors on the basis of their measured Eddington ratio,
contrary to the results of a number of studies of local AGN, which
have found evidence that high-accretion rate objects have smaller
$M_{\rm BH}/M_{*}$ ratio as compared to less active AGN (Greene and Ho
2006; Shen et al. 2008; Kim et al. 2008).

\subsection{Flow patterns in the $M_{\rm BH}$-$M_{*}$ plane}
As we have mentioned before, constraining the star formation rate of the AGN hosts in our sample is
a very difficult task, given the strong AGN component dominating the
SED in the rest-frame optical/UV bands. Individual estimates of SFR
based on the $\chi^2$ minimization procedure described above give
results which are uncertain by up to 0.7-0.8 dex. 
Nevertheless, despite these very large uncertainties on the estimated star formation
rates for our AGN hosts, we can try to assess the general direction of
{\it motion} of the objects in the $M_{\rm BH}-M_K$ plane; although
for each individual object it will be hard to accurately pin down the
change in total stellar mass, the overall ``ensemble'' average motion
of the flow could give interesting indications on the longer term
evolution of the scaling relations.

\begin{figure}[ht]
\epsscale{0.75}
\plotone{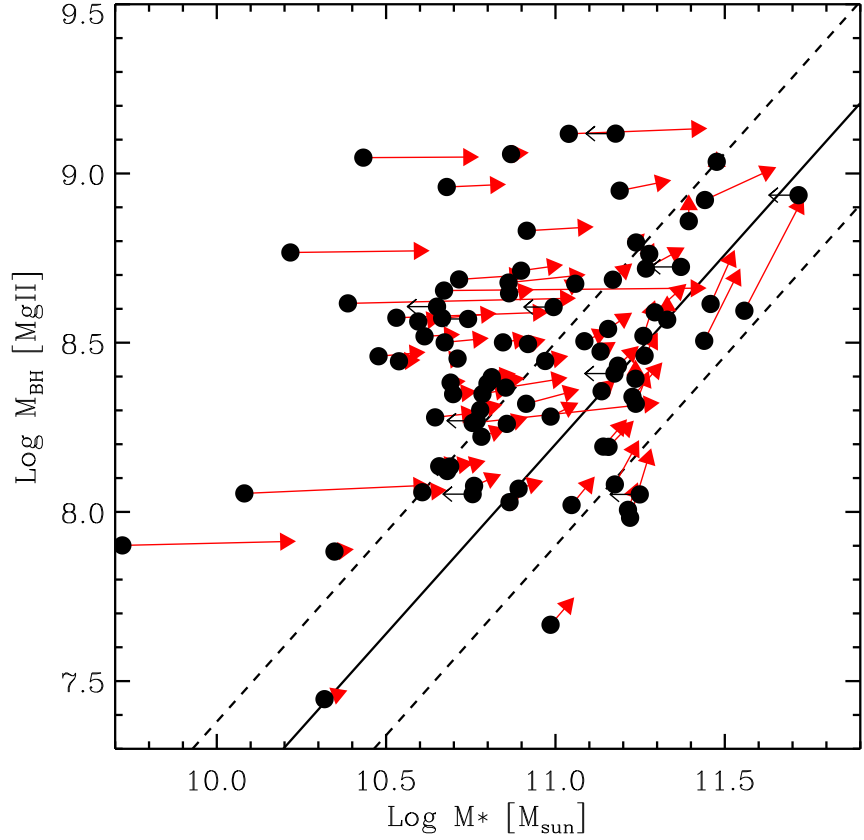}
\caption{Black hole mass host stellar mass relation for
  zCOSMOS type--1 AGN in the redshift range $1< z < 2.2$. 
Symbols with leftwards arrows represent upper limits on
the host mass. The black solid line is the best fit to the
\cite{haering:04} local spheroids sample relation, with dashed lines
marking a $\pm$0.3 dex offset.  Red
arrows represent the direction of evolution of the points in the
$M_{\rm BH}$-$M_{*}$ plane in 300 Myr on the basis of their instantaneous
accretion- and star formation-rates and an AGN duty-cycle estimated
from the amplitude of the corresponding luminosity and mass functions
(see text for details).} 
\label{fig:mbh_mk_mass_flow}
\end{figure}

To this end, we show in Fig.~\ref{fig:mbh_mk_mass_flow} as red thin
arrows the predicted flow patterns of the BLAGN. The tip of each arrow
marks the location where the system will find itself within $t_{\rm star}=300$ Myr, if
continually forming stars at the estimated star formation rates, while
at the same time the central black hole keeps accreting at the
measured rates for a fraction of this time equal to the AGN duty cycle
$\delta_t$. The exact value of $t_{\rm star}$ is of course arbitrary,
and has been chosen in order to better visualize the flow
pattern. Changing it, will simply rescale the length of the arrows,
but won't change their orientation.  
For each object observed at redshift $z$ having nuclear bolometric luminosity $L_{\rm bol}$ and
host galaxy stellar mass $M_{*}$, the duty cycle is estimated by
taking the ratio of the AGN bolometric luminosity function
\citep{hopkins:07} $\phi(L_{\rm bol},z)$ to the mass function of
highly star-forming galaxies \citep{ilbert:09}, which are assumed here
to represent the parent population of the sample at hand, $\phi_{\rm
  gal,HSF}(M_{*},z)$.
\begin{equation}
\delta_t(L_{\rm bol},M_{*},z)=\frac{\phi(L_{\rm bol},z)}{\phi_{\rm
  gal,HSF}(M_{*},z)}
\end{equation}
As the mass functions have a sharp exponential cut-off at high masses,
as opposed to the power-law decline of the QSO luminosity function,
the duty cycle defined above increase quickly for the most massive
hosts, causing the upwards turn of the flow pattern. 
Intriguingly, the predicted motion of the objects in the $M_{\rm
  BH}$-$M_{*}$ plane does lead to a reduced scatter of the points,
i.e. the flow appears to be ``converging''. To give a simple
quantitative estimate of this effect, we measure the scatter by
fitting all points (apart from those with only upper limits for the
host galaxy properties) with a linear relation. Given the reduced
dynamic range in black hole masses we are probing, and for the sake of
simplicity, we fix the slope of the correlation to the locally measured
one (1.12; H{\"a}ring and Rix 2004), and let the normalization be a
free parameter. The observed points have thus a normalization of about
$-3.7$ and an intrinsic scatter of 0.43. The tips of the arrows,
instead, move closer to the local relation (normalization $\approx
-3.9$) and with a much reduced scatter of 0.34.

Future observations of lower redshift AGN, for which an accurate
determination of both total stellar mass and star formation rate is
more easily achieved by a combination of multi-band photometry and
high-resolution spectroscopy will provide much better (and more
reliable) maps of the flow patterns of the AGN-host galaxy systems,
revealing fundamental details on their physical coupling.

\section{Systematics and selection effects}
\label{sec:bias}

The main result of the previous section is that our estimates of the
type--1 AGN host physical parameters are (although marginally) {\it inconsistent} with the
hypothesis that they lie on the $z=0$ scaling relation.
Here we wish to discuss how much of this observed offset can be due
to various systematics and selection effects.
We identify here two kind of biases: the first is the combination of
systematics inherent to our methods to measure either BH or host
galaxy's mass, and we will discuss these first (sections~\ref{sec:bias_imf},
\ref{sec:bias_bhmass} and \ref{sec:agn_template_heng}). 
The second, more subtle, is a 'luminosity/mass function weighted' bias on any AGN-selected
sample introduced by the intrinsic scatter in the scaling relations 
\citep{adelberger:05,fine:06,lauer:07,treu:07}. This
could induce a spurious effect on the measured offsets, provided the
scatter in the relation is large enough. We will 
discuss this effect in some detail in section~\ref{sec:sel_bias}.

\subsection{IMF and galaxy stellar masses}
\label{sec:bias_imf}
In this work, mainly to allow a direct comparison with most previous
works on the subject, we have adopted a Salpeter IMF to calculate the
total stellar mass of the host galaxies based on the BC03 SED fitting
procedure.
As mentioned already in section~\ref{sec:sed_fit}, adopting a
different IMF will result in a systematic shift of the best values for
$M_{*}$, typically reducing the stellar mass, and increasing the ratio
$M_{\rm BH}/M_{*}$. To quantify the systematic
uncertainty in the measured evolution introduced by the uncertainty in
the IMF, we have re-calculated the stellar masses using a Chabrier
IMF. We thus shifted the estimated values of the total stellar mass of
the AGN hosts by -0.255 dex (Pozzetti et al. 2007) 
and found the following values for the exponent of the redshift
evolution function: $\delta_2=1.15 \pm 0.13$. The smaller host masses
implied by the new choice of IMF result in a larger positive offset of
the $M_{\rm BH}/M_{*}$ ratio from the locally determined value, requiring
a more pronounced evolution. 
On the other hand, it should be
kept in mind that realistic SFH can be different from the smooth ones
adopted here, due to the presence of short bursts of star formation.
In fact, Pozzetti et al. (2007) have tested how the measured
stellar masses changes when random bursts are superimposed on smooth
SFH (with a similar range of combinations of $\tau$ and $t_{\rm age}$
to the one chosen here).
They conclude that, for $z>1.2$ samples, the mean mass computed with
smooth SFH is on average 0.16 dex smaller than the one computed using a bursty
SFH (see fig.5 in Pozzetti et al. 2007). 
This offset is smaller than our typical errors on the stellar
masses. Moreover, a robust assessment of the ''burstiness'' of the SFH
cannot be performed on our sample, as the number of extra parameters
required (duration of a burst, fraction of stellar mass produced in
the burst, time since last one) would surely introduce strong
degeneracies in our fits. 

Another well known source of systematic uncertainty lies in the choice
of the BC03 templates to describe the stellar populations of the AGN
hosts. Different groups have in recent years built different stellar
population models using different treatments of stellar structure and 
evolution (see e.g. Fioc \& Rocca-Volmerange (1997), Silva et
al. (1998), Maraston (2005)). A thorough discussion of the relative
differences among them is far beyond the scope and the aims of this
paper (but see, e.g., Longhetti and Saracco 2009). Suffice it to say,
in this context, that the typical offset in the estimated masses of
test galaxies with ages similar to those considered here are much
smaller (of the order of $\pm 0.1-0.15$ dex, see e.g. Cimatti et
al. 2008; Longhetti and Saracco 2009) 
than the statistical uncertainties of our mass measures.

Finally, it is clear that by comparing total stellar masses of AGN
hosts with the bulge/spheroid masses of the local galaxies originally used to
derive and identify the scaling relations, we are introducing a
significant bias (see also Jahnke et al. 2009; Bennert et al. 2009). 
Lacking any imaging information and reliable bulge-to-disc (B/T) decomposition, we can just argue that, at the very least,
the black hole to {\it bulge} mass ratio should show an even larger
offset from the local scaling relation. Systematic trends in the B/T
ratio with redshift (see e.g. Merloni, Rudnick and Di Matteo 2004)
will need also to be taken into account.

\subsection{Black hole mass measurements}
\label{sec:bias_bhmass}
We have already mentioned in section~\ref{sec:virial} that there is
currently a substantial uncertainty on the actual parameter of the
virial relationship to be used for the estimate of black hole masses
in broad line AGN. In the calculations so far, we have adopted the
McGill et al. (2008) expression, but different relationships, based on
different calibrations and/or assumptions about the broad line region
geometry exist in the literature.

To test the systematic effects on the measured evolution of the black
hole to host galaxy mass ratio evolution, we have re-calculated black
hole masses using the MLD04 and V09 relations (see
\S~\ref{sec:virial}). 

Adopting the relation of
McLure and Dunlop (2004), which has a steeper BLR size-luminosity
relation (0.62 instead of 0.5), substantially reduces the amount of
observed evolution: we obtain, for the exponents of the redshift
evolution function, $\delta_2= 0.47 \pm 0.12$, 
bringing it closer to the expectations of a purely
luminosity-bias dominated effect if the $M_{\rm BH}-M_{*}$ relation has
an intrinsic scatter as large as 0.5 dex (see section below).
On the other hand, adopting the V09 relation, which produce, on
average, larger black hole masses than in our fiducial case,  we
obtain $\delta_2=0.91 \pm 0.12$, 
indicating a larger amount of
positive evolution. It is worth keeping in mind, however, that the
McLure and Dunlop (2004) normalization of the ``virial'' black hole
mass estimate is based on a specific (theoretically motivated)
assumption on the BLR geometry, while the V09 one is empirically
calibrated with the local $M_{\rm BH}-\sigma$ relation, i.e. it does
not assume any preferred geometry.

Other possible systematic effects could be due to evolution of the
physical properties of the broad line region itself. However, no
significant trend with redshift was found in our data for either the
FeII emission strength (measured relative to the continuum
luminosity), or the overall goodness of the continuum fits, that could
have signalled an inadequacy in our model of the FeII template
(due, for example, to systematic changes in metallicity). Higher
signal-to-noise spectra are probably needed to assess these issues
with the due care.

Finally, we would like to note here that, when comparing with results
present in literature, one should take into account the different
methods of line fitting and \FeII\ subtraction. 
 A thorough analysis of the induced bias is not
straightforward, as the specific techniques of broad line fitting can
be different. Nevertheless, we have ourselves
performed a new fit of the \MgII\ line complex without including any
\FeII\ template in the continuum. 
The distribution of the (log of the) ratio between the
FWHM calculated without and with \FeII\ subtraction 
is centred at a positive offset of about 0.1 dex (which would correspond
to a difference in mass of about 0.2 dex), but with a significant tail
of higher ratios towards low values of the FWHM.

\subsection{Choice of AGN SED}
\label{sec:agn_template_heng}

In section~\ref{sec:sed_fit} we have shown how the measure of the
rest-frame K-band luminosity of the BLAGN hosts does not depend
strongly on the set of galaxy SED templates used to fit the composite
AGN+galaxy spectral energy distribution, provided that the same AGN
SED is used (we adopt here the Richards et al.~(2006) mean QSO SED,
allowing for additional dust extinction of the nuclear light). 

An obvious possible objection to our results is that, if the AGN SED
were markedly different from the Richards et al.~(2006) template at the
typical luminosities of the zCOMSOS bright sample, we would be
introducing a severe bias in our measures of the host galaxy
parameters. More worryingly, any systematic trend whereby AGN SEDs
change with either redshift, luminosity, black hole mass or Eddington
ratio, would introduce spurious trends in the measured evolution of the
offset from the local scaling relations.

Elvis et al. (2009, hereafter E09) performed a thorough, systematic study of the
spectral energy distribution of X-ray (XMM-COSMOS) selected,
spectroscopically confirmed, type--1 (broad line) AGN in the
COSMOS field. The interested reader is referred to E09
for a discussion of the main properties of the sample. Here we would
like to point out that, even when selecting only those objects that
are classified as ``pointlike'' from their ACS images, in the redshift
range of interest here ($1<z<2.2$) a
non-negligible contribution due to the compact stellar emission from the
host is still significantly present in the observed SED, even in
the brightest sources.

Indeed, the mean SED of the ``pointlike'' type--1 COSMOS AGN with
$L_{\rm bol}>10^{45.5}$ shows a less pronounced dip in the 1$\mu$m
region than either the Elvis et al.~(1994) or the Richards et
al.~(2006) templates. As a test, we have used this new, COSMOS based,
AGN SED together with the BC03 templates to fit the composite spectral
energy distribution of the objects in our sample. As expected, the
flatter mean AGN SED results in a {\it lower} residual host galaxy
contribution. The number of ``undetected'' hosts (i.e. they have only
upper limits in the rest frame K-band magnitude estimate) rises from
10 to 24; on average, the hosts
result about 0.4-0.5 magnitudes fainter. Consequently, the positive
offset of type--1 AGN from the local scaling relations would be
larger (we have measured, for these choice of AGN SED, $\delta_2\simeq
1.3$).

We argue that, on the basis of the photometric data only, it is not
possible to decide whether the apparent flatness of the AGN SED at NIR
wavelengths is indeed due to a dramatic change of intrinsic nuclear
continuum emission (thus minimizing the host galaxy contribution), or,
on the contrary, it is mainly due to stellar light contamination. As
the QSO SED we have adopted represent an extreme within the COSMOS
database (see E09), our approach, which {\it  maximizes} the host's
contribution is the most conservative one with respect to the measured
evolution of the $M_{\rm BH}/M_{*}$ ratio. Independent clues on the
intrinsic nuclear continuum shape from e.g. detailed spectroscopic
and/or polarimetric studies will be helpful.

Finally, we have also considered the effects of AGN variability on the
observed SED. This has been studied in detail by \cite{salvato:09},
where it was demonstrated that correcting for even relatively small variation of the
photometric points due to the non-simultaneous observation times can
significantly improve the photometric redshift determination for
AGN. A ``variability corrected'' photometric catalogue is available
for the subsample of X-ray detected type--1 AGN (82/89 objects in our
sample). We have thus recomputed the hosts total stellar masses and
studied the evolution of the $M_{\rm BH}/M_{*}$ ratio also for them,
finding no significant difference in the redshift dependence of the
measured offset ($\delta_2=0.84 \pm 0.13$).

\subsection{Selection bias}
\label{sec:sel_bias}
In this section we estimate the possible bias due to selection
effects.
Our objects are selected essentially on the basis of the nuclear (AGN)
luminosity, and on the detectability of the broad \MgII\ emission line,
clearly leading to a bias towards more massive black holes, similar to
Malmquist (1924) bias for luminosity selected samples of standard
candles.

In fact, there is a more subtle effect, generally applicable to all
cases where two properties of a class of objects are known to be
correlated with certain intrinsic dispersion, and one wishes to
determine the probability distribution of one of the two quantities,  
having selected objects on the basis of measurement of the other one.
For purely flux limited samples, this bias was already discussed and
calculated by Kellerman (1964, see Appendix), for the case in which the two
quantities were the spectral indices of AGN at two different radio
frequencies. This was then generalized by Francis (1993) to the case of
AGN spectral slopes in any two given independent bands.
For the specific case of black hole and host galaxy masses (or
velocity dispersion), this bias was discussed already in Adelberger and Steidel (2005),
but has been scrutinized in depth in Lauer et al. (2007), which also
recovered the Kellerman (1964) results for the specific case of a flux
limited sample, and we refer the curious reader to the Lauer et
al. (2007) paper for a more thorough discussion.

Here we make use of the main analytic results of Lauer et al. (2007)
and apply them to our particular selection criteria. The null
hypothesis we put under test is that the local scaling relation
between black hole and host galaxy mass, assumed here to be given by
eq.~(\ref{eq:haering04}), with an intrinsic scatter $\sigma_{\mu}$,
does not change with redshift, neither in normalization and slope, nor
in scatter.

\begin{figure}[ht]
\epsscale{0.75}
\plotone{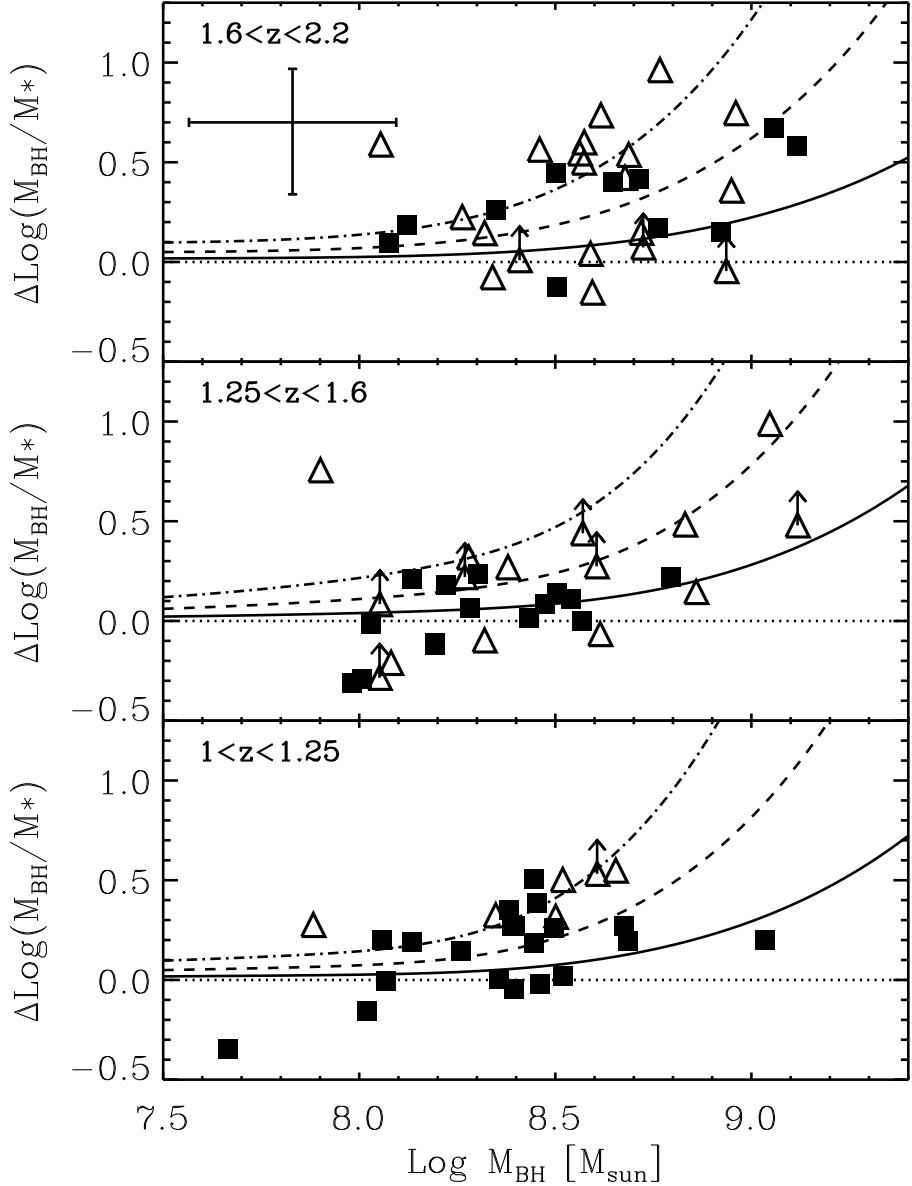}
\caption{Offset from the local $M_{\rm BH}-M_{*}$ relation as a function
  of black hole mass.  Filled
symbols represent measurements, 
symbols with upwards arrows represent upper limits on
the host mass.  Open triangles (filled squares) denote the objects
with low ($<0.33$) and high ($>0.33$)
Galaxy to AGN luminosity ratio in the rest frame K-band, respectively.
Lines show the bias due to the intrinsic
scatter to be expected even if the local relation is universal. Solid lines
are for an intrinsic scatter of 0.3 dex; dashed of 0.5 dex and
dot-dashed of 0.7 dex. The typical error bars of our measurements
are shown as black cross in the upper left corner.} 
\label{fig:dmbh_mbh}
\end{figure}

It is in fact the intrinsic scatter in the local relation, together
with the observed shape of the mass and/or luminosity function of the
selected objects in the appropriate redshift ranges that determine the
bias. In a nutshell, cosmic scatter in the $M_{\rm BH}-M_{*}$ relation implies that
there is a range of masses Log$M_{*}\pm \sigma_{\mu}$ for each object of a given black
hole mass $M_{\rm BH}$, where we have assumed, for simplicity, a
symmetric scatter in the relation. If the number density of galaxies is falling
off rapidly in the interval  Log$M_{*}\pm \sigma_{\mu}$, 
it will then be more likely to find one of the more numerous small
mass galaxies associated with the given black hole, and therefore a larger ratio $M_{\rm
  BH}/M_{*}$. Thus, given a distribution of galaxy masses (mass function
$\phi({\rm Log}M_{*})$), and provided that the scatter $\sigma_{\mu}$ is not too
large, the logarithmic offset of each point from the correlation
(\ref{eq:haering04}), assumed to be held fixed to the local
determination, is given by [cfr. Lauer et al. (2007), eq.(14)]:
\begin{equation}
\label{eq:bias_mbh}
\Delta{\rm Log}(M_{\rm BH}/M_{*}) = (1/S) \times \Delta{\rm Log}M_{\rm BH}
\approx \sigma^2_{\mu} \left(\frac{d {\rm Log} \phi}{d {\rm Log} M_{*}}\right)_{{\rm
    Log} M_{*}=(\mu-A)/B},
\end{equation}
where $1/S=0.67$ (see footnote \ref{foot:6} in section \ref{sec:scaling_evol}), $\mu={\rm Log}M_{\rm BH}$ and $(A,B)=(1.12,-4.12)$ are slope and
intercept of the relation (\ref{eq:haering04}).
We estimate the logarithmic derivative of the mass function $\frac{d
  {\rm Log} \phi}{d {\rm Log} M_{*}}$ using the mass function
  determination from the S-COSMOS galaxy survey (Ilbert et al. 2009;
  masses have been recalculated to adjust to our choice of Salpeter
  IMF) in the same redshift range probed by our
  BLAGN. Figure~\ref{fig:dmbh_mbh} shows our objects in the
  offset-black hole mass plane, where solid lines
  mark the expected bias from eq.~(\ref{eq:bias_mbh}) in three
  different redshift ranges, in the case of
  $\sigma_{\mu}=0.3$. Dashed and dot-dashed lines, instead, correspond to
  the cases of $\sigma_{\mu}=0.5$ and $0.7$, respectively. This plot
  clearly shows that the offset we measure is in excess to what
  expected in the most extreme case of large intrinsic scatter in
  the local relation, estimated by Novak et al. (2006) to be less than
  0.5 dex (see also G{\"u}telkin et al. 2009).
     This is mainly due to the fact that the
  AGN black hole masses we measure at the depth of the zCOSMOS selection function are not extremely
  large, and correspond, according to eq.~(\ref{eq:haering04}) to
  a range of host galaxy masses where the mass function is not falling
  off too steeply. On the other hand, our observations could be
  explained in terms of luminosity bias only if 
the scatter in the $M_{\rm BH}-M_{*}$ relation were
 as large as 0.5-0.7 at $z>1$.

Yet another test is possible, however. When studying the redshift evolution of $\Delta{\rm Log}(M_{\rm
  BH}/M_{*})$, fitting its redshift dependence with a functional
form (\S \ref{sec:scaling_evol}), one does effectively take the average of the
offset over a range of AGN luminosities above a well determined selection cut in any
given redshift bin. In this case, it can be shown (Lauer et al. 2007, eq.~25)
that the average offset at any given redshift is given by:
\begin{equation}
\langle \Delta{\rm Log}(M_{\rm BH}/M_{*}) \rangle (z) = \frac{0.67
  \sigma_{\mu}^2 [\Psi(L_{\rm min},z)-\Psi(L_{\rm max},z)]}{\int_{L_{\rm
      min}}^{L_{\rm max}}\Psi(L,z)dL},
\end{equation}
where $\Psi(L,z)$ is the type 1 AGN luminosity function at redshift
$z$ and $L_{\rm min}$ and $L_{\rm max}$ are the minimum and maximum
luminosity of the AGN that can enter our sample at the same redshift,
given our survey selection function. We have calculated this bias
adopting the VVDS type 1 AGN luminosity function of Bongiorno et
al.~(2007), and the results are plotted as red 
 lines in Figure~\ref{fig:dmbh_z_mass}. For our type--1 AGN
sample, which extends well below the knee of the type 1 luminosity
function in this redshift range, the expected bias in the offset is
almost constant with redshift, amounting to $\approx 0.3$ (0.1) dex
for $\sigma_{\mu}=0.5$ (0.3). This is consistent with the estimate for
$\Delta{\rm Log}M_{\rm BH}$ of Lauer et al. (2007) based on the local
AGN luminosity function of \cite{boyle:00}, and slightly smaller than the
measured offset. On the other hand, a larger intrinsic scatter in the
scaling relation at the redshift of interest (up to $\sigma_{\mu}=0.7$,
dotted lines in Fig.~\ref{fig:dmbh_z_mass}) could indeed be the
cause of the measured offset in our sample.

A note of caution is however in place. Our measured offset
 is based on a black hole mass ``virial'' estimator
that has been implicitly calibrated on the local $M_{\rm BH} -
\sigma_{*}$ relation \citep{onken:04}. This is usually done with the
justification that the true geometry of the broad line region is not
known, which enters as a multiplicative factor in the virial relation.
However, the reverberation-mapped AGN used by Onken et al. (2004) to
carry out this normalization, could themselves be affected by the
'luminosity/mass function weighted bias'. If that were the case, and if the local
AGN samples spanned a similar mass and luminosity range as ours, the
bias would be much reduced, and possibly canceled out completely (as
suggested also by the results of Kim et al. 2008). The
selection function of the Onken et al. (2004) AGN is, however, far from
being well understood, and a quantitative estimate of the true
expected residual bias is beyond the scope of this work.
As we discussed above, large, uniformly selected samples of AGN hosts, spanning a larger
redshift range than probed here,
as well as accurate comparisons of scaling relations for active
(reverberation mapped) and
non-active galaxies at low-$z$ with similar selection functions, are and
will be very important in disentangling
true redshift evolution from other biases and systematic differences with the
local samples.

Finally, we would like to mention the further aspect of what, in
general terms, can be defined luminosity bias of the AGN, namely the
fact that faint broad line AGN cannot be 
detected in bright galaxies. This is clearly 
an important issue to consider in our case,
given this zCOSMOS sample has a larger fraction of faint AGNs. 

In order to assess this, we have divided sources in Fig.~\ref{fig:mbh_mk_mass}
and Fig.~\ref{fig:dmbh_mbh} according to the measured luminosity ratio in the
rest-frame K-band, with open triangles (filled squares) marking the
most (least) AGN-dominated sources. It is interesting here to point out that,
indeed, high contrast (AGN-dominated) objects do seem to slightly 
bias the result toward high $M_{\rm BH}/M_{*}$ ratios. However,
removing all such objects, the main results presented here of a redshift
evolution of the scaling relation, are not
only confirmed, but do appear to be strengthened, 
as AGN-dominated objects (open triangles) are the largest outliers
also in the lowest redshift bin.

\section{Discussion and conclusions}
\label{sec:disc}

\subsection{Our results}
\label{sec:our_results}
We have used an AGN+host galaxy SED decomposition technique to infer
the physical properties of the hosts of 89 (moderately luminous:
i.e. mostly sub-$L*$) 
type--1 AGN in the zCOSMOS survey.
Thanks to the deep, intensive multi-wavelength coverage of the COSMOS
field, the observed spectral energy distributions are sampled to such
a degree that the decomposition technique works reasonably well. 
We are thus able to derive rest frame
    K-band magnitudes and total stellar masses for the majority of our
    sample (80-90\%, depending on the choice of AGN SED). 
    Noticeably, our method allows us to properly quantify the uncertainty in
    each measurement.

The bulk of the sample of BLAGN hosts we have studied have total
    stellar masses in the range 10$^{10.5}$--10$^{11.3}
    M_{\odot}$. Both the derived mass-to-light ratios and the sample
    average star formation rates seem to suggest that they
    are moderately-to-highly star forming objects, in good agreement with the host
properties of both type 1 and type 2 (obscured) X-ray and/or IR
selected AGN, either in a
lower (Kauffmann et al. 2003; Jahnke et al. 2004; Hickox et al. 2009; Silverman et al. 2009)
and in a similar (Brusa et al. 2009; Ross et al. 2009) redshift
range. Reassuringly,
also the fits with the P07 phenomenological templates indicate an
overall preference for star-forming hosts over passive (elliptical)
ones. Obviously, a more
    accurate determination of the individual star formation rates is
    hampered by the dominant contribution of the AGN in the rest-frame
    UV/Optical part of the spectra. 
 
The black hole masses derived with the the ``virial'' or
``empirically calibrated photo-ionization'' method are broadly
distributed around Log$M_{\rm BH}\sim 8.5$. Interestingly, the mass
range probed by our sample corresponds to that where the most
stringent constraints on the $z=0$ scaling relations are available.
The Eddington ratios $\lambda$ of our objects have a mean 
of Log$\lambda\approx -1$. We observe a clear
    trend between Eddington ratio and bolometric luminosity,
    that could be indicative of some specific luminosity-dependent AGN
  lifetimes distribution \citep{hopkins:09b}, but the underlying effect of various
  selection biases in determining this trend needs to be further
  assessed \citep{trump:09}.

For a large number of objects in the sample (68/89; but the number
falls to 24/89 if we consider the errors) the measured black
    hole to stellar host mass ratio is positively offset from that
    predicted by the local H{\"a}ring \& Rix (2004) local scaling relation. Assuming a
    redshift-dependent evolution in the $\Delta{\rm Log}(M_{\rm BH}/M_{*})$ of the
    form $\delta_2{\rm Log}(1+z)$, we measure $\delta_2=0.68 \pm 0.12
    ^{+0.6}_{-0.3}$, where the large asymmetric systematic
errors stem from the uncertainties in the hosts' IMF, in the
calibration of the virial relation used to estimate BH masses and in
the mean QSO SED to be used. 

The scatter in the measured offset is substantial at all redshifts
probed, such that we could still be consistent with a lack of
evolution in the scaling relation at the 2-$\sigma$ level, as
shown by the inset of figure~\ref{fig:mbh_mk_mass}. There, it is also
apparent that the majority of observational
data-points for AGN samples in the published 
literature do indeed show a broadly consistent amount
of offset at all redshifts probed. This might suggest that intrinsic
differences in the SMBH/host galaxy relation between active and
inactive galaxies could play an important role besides any genuine
cosmological evolution.

Jahnke et al. (2009) have independently computed total stellar masses
for a small sample (18 objects) of X-ray selected AGN in the COSMOS field for
which simultaneous {\it HST}/ACS and {\it HST}/NICMOS observations
allow to clearly image the resolved hosts. 
The two independent methods for measuring stellar masses are in broad
agreement with each other, with a dispersion well within the (large)
uncertainties that characterize each of these methods. 
5/18 of them have also zCOSMOS spectra, and
are part of the sample described here (empty stars in
Fig.~\ref{fig:mbh_mk_mass}). 
However, the objects in the Jahnke et al. (2009) sample do not show
any significant offset from the local scaling relation. This might be
due to a statistical fluctuation (more so given that they tend to lie in
the lower redshift range probed by our sample, where the offset from
the local scaling relation is smaller), or may indicate a more serious
issue with the different estimates of the mass-to-light ratio of the AGN hosts. 
Larger samples of {\it HST}-imaged AGN hosts at longer wavelengths
(with the newly installed WFC3) will be extremely important to settle
this issue and, in general, to improve the calibration of our
SED-based method to estimated host galaxy masses for AGN.

We have taken particular care in examining the effects of the
    bias inevitably introduced by any intrinsic scatter in the BH-host
  mass scaling relation into any AGN-selected sample, as ours.
We conclude that our data cannot possibly be explained if
       type--1 AGN and their hosts at $1<z<2$ lie on a scaling
       relation which has the same slope, normalization and scatter as
     the locally observed one. On the other hand positive evolution of
   the average $M_{\rm BH}/M_{*}$ ratio (larger black holes at early
   times in unobscured AGN) or of the intrinsic scatter (or a
   combination of the two) are needed to explain our results.

\subsection{Implications for theoretical models}
What are the implications of these findings for our understanding of
the cosmological co-evolution of black holes and galaxies?
Let us briefly discuss recent theoretical investigations on this issue
and the corresponding predictions for the evolution of the scaling
relations.

One of the earliest semi-analytic models to incorporate the evolution
of supermassive black holes and the associated feedback effect were
those by Granato et al. (2001,2004). There, triggering of AGN activity
is not directly linked to merger activity, but rather generically to
the process of bulge/spheroid formation. The rate of star formation
and black hole accretion is regulated by the starlight radiation drag,
and consequently, in a typical system the ratio $M_{\rm BH}/M_{*}$ is
initially small and rapidly grows until the AGN feedback sweeps the
remaining gas. QSOs and, in general, type--1 AGN are thus associated
to the final stage of bulge formation, and it is very hard to produce
any positive offset from the local relation like the one we measure.

This is however a problem common to all feedback models in which the
black hole energy injection is very fast (explosive). Indeed, the
first published predictions of merger-induced AGN activity models
\citep{robertson:06} indicated that, if strong QSO feedback is
responsible for rapidly terminating star formation in the bulge
\citep{dimatteo:05,springel:05}, then very little evolution, as well
as very little scatter, is expected for the scaling relations.
However, later works within the same theoretical framework \citep{hopkins:07b,hopkins:09a}
have analyzed in greater depths the role of dissipation in major
mergers at different redshift. Under the assumption that black hole
and spheroids obey a universal ``black hole fundamental plane''
(BHFP), where $M_{\rm BH}\propto M_{*}\sigma_{*}^2$, they show
how, in
gas-richer environments (at higher redshift), dissipation effects may
deepen the potential well around the black hole, allowing it to grow
above the $z=0$ $M_{\rm BH}-M_{*}$ relation, to a degree marginally
consistent with our results. However, in the same physical framework,
the $M_{\rm BH}-\sigma_{*}$ relation is almost independent on
redshift, which would contradict the observational results of
\cite{woo:06,treu:07,woo:08}. 

Another, related effect was discussed in Croton (2006). There it was
assumed that major mergers can trigger both star formation in a bulge
as well as black hole growth, in a fixed proportion. However,
bulges can also acquire mass by disrupting stellar discs, a channel
that should not contribute to black hole growth. The relative
importance of these two paths of bulge formation may lead to lighter
bulges for a given black hole mass at high redshift, as disks have a
smaller stellar fraction. A subsequent
study of this and other dynamical process of disk-to-bulge
transformation was included in the work by Fontanot et al.~(2006) and 
Malbon et al.~(2007). They
also confirmed qualitatively the predictions of Croton (2006), but
found a much smaller effect, at most a factor $\sim 2$ at $z=2$ in the
Malbon et al. (2007) work, and
preferentially for small mass black holes $M_{\rm BH} \la 10^{8} M_{\odot}$.
Even more complex semi-analytic models including various flavors of
AGN-driven winds and their feedback effects \citep{fontanot:06} can
lead to various degrees of positive redshift evolution of the average
$M_{\rm BH}/M_{*}$ ratio (Lamastra et al. 2009).

As a general rule, we observe that, following increasing observational
efforts to study the evolution of scaling relations, semi-analytic
models have become more sophisticated over the years. This increase of
sophistication's has allowed more complex behaviours of the coupled
black holes-galaxy systems over cosmological times. As it is expected,
more complex models also lead to an increase in the predicted scatter,
even though a clear theoretical study on the redshift evolution of
such scatter is still missing (but see the recent attempts by Lamastra
et al. 2009 and Somerville 2009).

Hints from hydrodynamical simulations, both of isolated mergers
\citep{johansson:09} and of relatively small cosmological boxes
\citep{coldberg:08} do indeed show a large scatter in the
instantaneous ratio between black hole accretion and star formation
rates, similar to what was found here (see also Silverman et
al. 2009), thus suggesting that on the relatively short timescales over
which un-absorbed AGN/QSOs are visible the physical connection between
black holes growth and galaxy formation must be complex, too. How this
would impact on the statistical and evolutionary properties of the
galaxy population as a whole, however, is far from clear. 

The results we have presented in \S \ref{sec:scaling_evol}, coupled with 
analysis of selection biases of \S \ref{sec:sel_bias} would suggest
that a greater effort should be made by theoretical modellers to
include a more accurate and realistic study of the evolution of the
intrinsic scatter in any scaling relations, as well as that of slope
and normalization.

The ``increased scatter'' hypothesis, as an explanation of the
observed offset, and its inevitable consequence
that a 'luminosity function weighted' bias plays a significant role in the observed evolution
of scaling relations, could be strengthened if
    recent claims of under-massive black holes in IR selected galaxy
    samples were confirmed \citep{shapiro:09}. Indeed, we should expect that in
    samples selected purely on the basis of the host galaxy stellar mass,
    rather than on AGN properties, the intrinsic scatter in the
    scaling relation should produce a bias going in the opposite
    direction as those discussed above, depending on the exact shape
    of the BH mass function at the redshift considered.

\subsection{Concluding remarks}
By taking advantage of the unique combination of VLT spectroscopy and
deep multi-wavelength coverage of the COSMOS field, we have presented
here a novel method to study the physical link between supermassive
black holes and their host galaxies in type--1 (un-obscured) AGN in
the crucial redshift range $1\la z \la 2$. The main focus of this work
is on the capability of our SED decomposition technique to provide
reliable estimates of the total stellar mass of the AGN hosts, and,
even more importantly, reliable estimates of its uncertainty.

The main result of our study is the observation of an offset in
the $M_{\rm BH}-M_{*}$ relation, such that, in the redshift range
probed, for their given
hosts black holes are on average 2-3 times larger than their
counterparts in the nuclei of nearby inactive galaxies.
A thorough analysis of all possible observational biases induced by
intrinsic scatter in the scaling relations reinforces the conclusion
that an evolution of the $M_{\rm BH}-M_{*}$ relation must
ensue for actively growing black holes at early times:
either its overall normalization, or its intrinsic scatter (or
both) must increase significantly with redshift. 

We close with two recommendations for future studies of the subject.
From the observational point of view, it will be very important to
explore methods to derive robust black hole mass estimates in high
redshift samples of obscured AGN, that can be 
selected purely on the basis of their host galaxy
properties. Broad emission lines at longer wavelengths, where the
effect of obscuration are less severe, could be very useful in this
respect. Also, a better understanding of the differences in the
hosts' properties of active and inactive black holes is needed to
allow a more meaningful comparison with the local scaling relations,
and a better assessment of their evolution.
From the theoretical point of view, more efforts should be devoted to
derive robust predictions for the coupled
evolution of slope, normalization {\it and intrinsic scatter} in the scaling
relations, and to properly include in the models the selection effects
that clearly play an often decisive role in the observational studies
of co-evolving galaxies and black holes.

 
 \acknowledgments
This research was supported by the DFG  cluster of excellence 'Origin
and Structure of the Universe'. We thank the anonymous referee for
carefully reading the manuscript and providing us with insightful and
constructive remarks. We also thank R. Decarli, F. Fontanot,
O. Gerhard, J. Greiner,
P. Hopkins, S. Komossa, N. Menci, P. Monaco, H. Netzer, R. Saglia, 
L. Wisotzki, J.-H. Woo for their 
useful comments and suggestions.
 The HST COSMOS Treasury program was supported through NASA grant
HST-GO-09822. We wish to thank Tony Roman, Denise Taylor, and David 
 Soderblom for their assistance in planning and scheduling of the extensive COSMOS 
 observations.
 We gratefully acknowledge the contributions of the entire COSMOS collaboration
 consisting of more than 80 scientists. 
 More information on the COSMOS survey is available at
  {\tt http://www.astro.caltech.edu/$\sim$cosmos}. It is a pleasure the 
 acknowledge the excellent services provided by the NASA IPAC/IRSA 
 staff (Anastasia Laity, Anastasia Alexov, Bruce Berriman and John Good) 
 in providing online archive and server capabilities for the COSMOS datasets.
 In Italy this work was partly supported by an INAF contract
 PRIN/2007/1.06.10.08 and an ASI grant ASI/COFIS/WP3110 I/026/07/0; in
 Mexico by the CONACyT grant program 83564 and PAPIIT IN110209; in
 Germany  by the Bundesministerium f{\"u}r Bildung und 
Forschung/Deutsches Zentrum f{\"u}r Luft und Raumfahrt and the 
Max Planck Society.

 \clearpage

 \clearpage
 
 \begin{deluxetable}{lccccccc}
 \tabletypesize{\scriptsize}
 \tablecaption{Rest frame K-band Magnitude and total stellar masses 
of BLAGN hosts in COSMOS \label{tab:bc}}
 \tablewidth{0pt}
 \tablehead{
 \colhead{VIMOS ID} & RA (deg) & DEC (deg) & \colhead{$z$} &
 \colhead{$M_{K}$[Vega]} & \colhead{$f_{\rm gal,K}$\tablenotemark{a}} &
 \colhead{Log $M_{*}$ [$M_{\odot}$]} & \colhead{ul\tablenotemark{b}} } 
 \startdata
 801709  &  149.985992 &  1.617284  &  1.126  & -25.24  &  0.847  &  11.06  & 0  \\
 803695  &  150.596069 &  1.787450  &  1.246  & -24.81  &  0.281  &  10.70  & 0  \\
 805949  &  150.050400 &  1.744427  &  1.149  & -25.74  &  1.459  &  10.99  & 0  \\
 807560  &  149.687286 &  1.719174  &  1.349  & -24.27  &  0.077  &  10.43  & 0  \\
 808150  &  149.545349 &  1.668093  &  2.096  & -26.88  &  0.446  &  11.44  & 0  \\
 810061  &  150.536713 &  1.849565  &  1.824  & -24.35  &  0.142  &  10.48  & 0  \\
 811239  &  150.278976 &  1.959607  &  1.550  & -26.15  &  0.307  &  11.18  & 0  \\
 811646  &  150.195053 &  1.793735  &  1.851  & -26.35  &  0.317  &  10.59  & 0  \\
 811799  &  150.161789 &  1.877919  &  1.444  & -24.68  &  0.536  &  10.66  & 0  \\
 811904  &  150.141220 &  1.819711  &  1.192  & -25.01  &  0.310  &  10.35  & 0  \\
 811960  &  150.131744 &  1.799389  &  1.675  & -25.53  &  0.372  &  10.90  & 0  \\
 813283  &  149.821960 &  1.838634  &  1.351  & -25.96  &  0.408  &  11.16  & 0  \\
 813416  &  149.791794 &  1.872849  &  1.567  & -25.77  &  0.596  &  11.21  & 0  \\
 813886  &  149.687500 &  1.812649  &  1.215  & -24.45  &  0.596  &  10.69  & 0  \\
 814414  &  149.564636 &  1.823087  &  1.507  & -26.22  &  0.170  &  11.25  & 1  \\
 816818  &  150.446045 &  2.043490  &  1.171  & -25.30  &  0.391  &  10.92  & 0  \\
 817202  &  150.386765 &  1.966629  &  1.537  & -25.62  &  0.415  &  10.99  & 0  \\
 817260  &  150.373642 &  2.112055  &  1.914  & -25.02  &  0.096  &  10.68  & 0  \\
 817480  &  150.328293 &  2.124951  &  1.780  & -25.78  &  0.425  &  10.87  & 0  \\
 818094  &  150.195587 &  2.004415  &  1.923  & -26.52  &  0.182  &  11.37  & 1  \\
 819187  &  149.957733 &  2.003069  &  1.806  & -25.12  &  0.168  &  10.72  & 0  \\
 819193  &  149.955856 &  2.028046  &  1.756  & -27.64  &  0.579  &  11.44  & 0  \\
 819446  &  149.897934 &  2.093906  &  1.910  & -25.11  &  0.377  &  10.68  & 0  \\
 819579  &  149.868561 &  1.992970  &  1.166  & -24.85  &  0.523  &  10.69  & 0  \\
 819592  &  149.865585 &  2.003061  &  1.248  & -25.44  &  0.742  &  10.89  & 0  \\
 819644  &  149.851959 &  1.998422  &  1.244  & -26.88  &  0.326  &  10.67  & 0  \\
 819702  &  149.837067 &  2.008842  &  1.481  & -24.82  &  0.143  &  10.64  & 0  \\
 820341  &  149.663605 &  2.085205  &  1.220  & -25.10  &  0.273  &  10.85  & 0  \\
 820375  &  149.656326 &  2.051113  &  1.855  & -25.14  &  0.586  &  10.78  & 0  \\
 820673  &  149.586609 &  2.037102  &  1.850  & -25.04  &  0.149  &  10.67  & 0  \\
 820679  &  149.585220 &  2.051113  &  1.355  & -25.54  &  0.137  &  10.92  & 0  \\
 821039  &  149.506729 &  2.074688  &  1.226  & -25.55  &  0.571  &  11.24  & 0  \\
 821885  &  150.708008 &  2.292316  &  1.099  & -25.78  &  0.607  &  11.26  & 0  \\
 822461  &  150.581863 &  2.287697  &  1.343  & -25.89  &  0.239  &  11.39  & 0  \\
 822703  &  150.536163 &  2.273239  &  1.087  & -25.61  &  1.854  &  10.97  & 0  \\
 823199  &  150.451859 &  2.144812  &  1.298  & -25.00  &  0.206  &  10.80  & 0  \\
 823714  &  150.345932 &  2.147529  &  1.258  & -26.16  &  0.841  &  11.22  & 0  \\
 824176  &  150.236267 &  2.289114  &  2.078  & -25.11  &  0.147  &  10.22  & 0  \\
 824306  &  150.214676 &  2.204261  &  1.841  & -24.03  &  0.068  &  10.08  & 0  \\
 824390  &  150.199768 &  2.190844  &  1.510  & -25.79  &  0.381  &  11.09  & 0  \\
 824396  &  150.198990 &  2.132499  &  2.160  & -27.12  &  0.223  &  11.56  & 0  \\
 824572  &  150.158371 &  2.139555  &  1.828  & -26.61  &  0.207  &  11.29  & 0  \\
 825363  &  150.004471 &  2.237096  &  1.407  & -25.77  &  0.317  &  11.24  & 0  \\
 825899  &  149.895416 &  2.239492  &  1.742  & -25.52  &  0.179  &  10.91  & 0  \\
 825906  &  149.894852 &  2.174454  &  1.323  & -25.48  &  0.991  &  11.13  & 0  \\
 827274  &  149.624283 &  2.180656  &  1.185  & -25.93  &  0.478  &  11.05  & 0  \\
 829667  &  150.499130 &  2.444901  &  2.025  & -27.36  &  0.560  &  11.04  & 0  \\
 829682  &  150.495667 &  2.412547  &  1.370  & -26.13  &  0.818  &  11.18  & 0  \\
 830510  &  150.347702 &  2.390998  &  1.848  & -26.02  &  0.379  &  11.28  & 0  \\
 831077  &  150.231812 &  2.363971  &  1.936  & -25.02  &  0.093  &  10.53  & 0  \\
 832354  &  149.993912 &  2.301415  &  1.789  & -25.56  &  0.130  &  10.86  & 0  \\
 832715  &  149.919785 &  2.327419  &  1.454  & -26.04  &  0.136  &  11.18  & 1  \\
 832923  &  149.881012 &  2.450839  &  1.315  & -26.16  &  0.957  &  11.14  & 0  \\
 832961  &  149.871841 &  2.342855  &  1.735  & -25.19  &  0.487  &  10.76  & 0  \\
 832963  &  149.870712 &  2.417283  &  1.528  & -25.06  &  0.353  &  10.78  & 0  \\
 833273  &  149.812943 &  2.345459  &  1.800  & -26.64  &  0.464  &  10.67  & 0  \\
 833541  &  149.763458 &  2.334125  &  1.131  & -25.24  &  0.431  &  10.81  & 0  \\
 833712  &  149.730255 &  2.453799  &  1.101  & -24.58  &  1.592  &  10.61  & 0  \\
 833817  &  149.705872 &  2.419752  &  1.108  & -26.16  &  0.525  &  11.17  & 0  \\
 834079  &  149.660278 &  2.410915  &  1.161  & -26.28  &  0.768  &  11.48  & 0  \\
 834383  &  149.602066 &  2.392675  &  1.849  & -27.39  &  0.215  &  11.72  & 1  \\
 834988  &  149.462845 &  2.356840  &  1.185  & -24.72  &  0.186  &  10.65  & 1  \\
 835006  &  149.459167 &  2.430080  &  1.242  & -23.80  &  0.339  &  10.32  & 0  \\
 835631  &  150.715118 &  2.484831  &  1.996  & -26.44  &  0.117  &  10.76  & 0  \\
 835840  &  150.668900 &  2.516766  &  1.573  & -23.39  &  0.088  &   9.72  & 0  \\
 836198  &  150.597351 &  2.617924  &  1.447  & -26.12  &  1.295  &  11.16  & 0  \\
 836355  &  150.572617 &  2.499909  &  1.102  & -24.70  &  0.190  &  10.61  & 0  \\
 837652  &  150.334427 &  2.561485  &  1.832  & -26.31  &  0.234  &  11.19  & 0  \\
 837827  &  150.305588 &  2.602232  &  1.342  & -24.98  &  0.188  &  10.76  & 1  \\
 837858  &  150.299744 &  2.506903  &  1.506  & -25.01  &  0.222  &  10.77  & 1  \\
 838223  &  150.230820 &  2.578165  &  1.401  & -25.58  &  0.132  &  10.99  & 1  \\
 838610  &  150.163803 &  2.597661  &  1.589  & -25.09  &  0.367  &  10.78  & 0  \\
 839188  &  150.058716 &  2.477386  &  1.256  & -24.95  &  0.095  &  10.74  & 1  \\
 839751  &  149.955612 &  2.502021  &  1.458  & -25.85  &  0.583  &  11.24  & 0  \\
 841635  &  149.575638 &  2.575658  &  1.171  & -24.50  &  0.590  &  10.54  & 0  \\
 843302  &  150.635406 &  2.649920  &  1.223  & -25.22  &  0.343  &  10.86  & 0  \\
 843685  &  150.554871 &  2.641009  &  1.144  & -25.36  &  0.503  &  10.81  & 0  \\
 844213  &  150.456650 &  2.648144  &  2.050  & -26.43  &  0.279  &  11.27  & 0  \\
 845220  &  150.251328 &  2.737164  &  2.162  & -25.32  &  0.104  &  10.85  & 0  \\
 845272  &  150.240799 &  2.659021  &  1.410  & -24.87  &  0.449  &  10.86  & 0  \\
 845728  &  150.147079 &  2.717479  &  1.177  & -25.62  &  1.152  &  11.14  & 0  \\
 845970  &  150.104462 &  2.691239  &  1.882  & -26.03  &  0.272  &  11.17  & 1  \\
 846335  &  150.042480 &  2.629174  &  1.569  & -26.63  &  0.251  &  11.46  & 0  \\
 847623  &  149.774170 &  2.674153  &  1.108  & -26.02  &  1.851  &  11.26  & 0  \\
 848220  &  149.621902 &  2.738307  &  1.889  & -25.78  &  0.088  &  10.39  & 0  \\
 851007  &  150.158981 &  2.825123  &  1.856  & -26.16  &  0.332  &  11.23  & 0  \\
 900028  &  150.084366 &  2.290529  &  1.112  & -24.79  &  0.595  &  10.71  & 0  \\
 900066  &  149.551071 &  2.316902  &  1.432  & -25.60  &  0.462  &  11.33  & 0  \\
 950021  &  149.574463 &  2.085072  &  1.623  & -25.38  &  0.452  &  10.86  & 0  \\

 \enddata
\tablenotetext{a}{Ratio of galaxy to AGN luminosity in the rest frame
  K-band, as derived from the SED decomposition technique.}
 \tablenotetext{b}{Upper limit flag: 1=upper limit}
  \end{deluxetable}

 \clearpage
 \begin{deluxetable}{lccccccc}
 \tabletypesize{\scriptsize}
 \tablecaption{\MgII\ broad emission line FWHM, continuum luminosity
   and black hole mass for zCOSMOS type--1 AGN \label{tab:agn}}
 \tablewidth{0pt}
 \tablehead{
 \colhead{VIMOS ID} & \colhead{$m_i$\tablenotemark{a}} & \colhead{FWHM$_{1000}$\tablenotemark{b}} &  \colhead{$\Delta$FWHM} &
 \colhead{$ L_{\rm 3000,40}$\tablenotemark{c}} &  \colhead{$ \Delta
   L_{\rm 3000,40}$} & \colhead{Log $M_{\rm BH}$
   [$M_{\odot}$]\tablenotemark{d}} & Radio Flag\tablenotemark{e}}
\startdata
 801709  &  22.41  &   9.23\tablenotemark{f}  &   0.19  &   2.97  &   0.15  &   8.67  & RQ  \\
 803695  &  21.39  &   4.98  &   0.06  &   8.34  &   0.20  &   8.35  & RQ  \\
 805949  &  21.74  &   2.52  &   0.15  &   5.39  &   0.21  &   7.67  & RL  \\
 807560  &  20.36  &   7.82  &   0.21  &  37.48  &   0.48  &   9.05  & RQ  \\
 808150  &  20.20  &   3.45  &   0.11  &  85.33  &   5.14  &   8.51  & RQ  \\
 810061  &  21.71  &   4.69  &   0.41  &  18.64  &   1.10  &   8.46  & RQ  \\
 811239  &  20.83  &   2.63  &   0.05  &  33.89  &   0.92  &   8.08  & RQ  \\
 811646  &  21.99  &   5.48  &   0.41  &  15.78  &   1.51  &   8.56  & RL  \\
 811799  &  22.41  &   3.80  &   0.29  &   9.21  &   0.37  &   8.13  & RQ  \\
 811904  &  22.22  &   3.74  &   0.56  &   2.89  &   0.18  &   7.88  & RQ  \\
 811960  &  20.90  &   5.46  &   0.81  &  33.60  &   0.97  &   8.71  & RQ  \\
 813283  &  20.88  &   4.97  &   0.21  &  21.60  &   0.30  &   8.54  & RQ  \\
 813416  &  22.42  &   3.63  &   0.25  &   5.98  &   0.56  &   8.01  & RQ  \\
 813886  &  22.42  &   6.49  &   0.28  &   3.18  &   0.21  &   8.38  & RQ  \\
 814414  &  20.34  &   2.70  &   0.12  &  26.15  &   0.39  &   8.05  & RQ  \\
 816818  &  21.14  &   5.46  &   0.07  &  11.64  &   0.27  &   8.50  & RQ  \\
 817202  &  21.27  &   3.80  &   0.10  &  18.89  &   0.54  &   8.28  & RQ  \\
 817260  &  20.95  &   6.77  &   0.09  &  45.04  &   1.80  &   8.96  & RQ  \\
 817480  &  22.30  &  10.46  &   0.26  &  11.40  &   1.59  &   9.06  & RQ  \\
 818094  &  20.13  &   4.42  &   0.17  &  87.15  &   2.41  &   8.72  & RL  \\
 819187  &  21.03  &   5.29  &   0.06  &  33.82  &   1.63  &   8.69  & RQ  \\
 819193  &  19.00  &   4.46  &   0.21  & 221.60  &   2.12  &   8.92  & RQ  \\
 819446  &  21.73  &   3.35  &   0.15  &  14.75  &   1.53  &   8.12  & RQ  \\
 819579  &  21.69  &   3.95  &   0.14  &   7.83  &   0.27  &   8.13  & RQ  \\
 819592  &  21.90  &   3.82  &   0.11  &   6.55  &   0.21  &   8.07  & RQ  \\
 819644  &  18.53  &   3.65  &   0.45  & 139.86  &   6.38  &   8.65  & RQ  \\
 819702  &  20.93  &   3.56  &   0.16  &  24.83  &   0.63  &   8.28  & RQ  \\
 820341  &  20.62  &   4.69  &   0.58  &  22.64  &   0.31  &   8.50  & RQ  \\
 820375  &  22.45  &   5.02  &   0.16  &   8.07  &   0.77  &   8.35  & RQ  \\
 820673  &  21.06  &   4.97  &   0.07  &  25.19  &   1.34  &   8.57  & RQ  \\
 820679  &  19.82  &   5.39  &   0.07  &  63.30  &   0.65  &   8.83  & RQ  \\
 821039  &  22.01  &   5.97  &   0.43  &   4.79  &   0.20  &   8.39  & RQ  \\
 821885  &  20.99  &   5.48  &   0.19  &  12.85  &   0.22  &   8.52  & RQ  \\
 822461  &  21.29  &   7.93  &   0.52  &  14.01  &   0.32  &   8.86  & RQ  \\
 822703  &  21.58  &   5.88  &   0.31  &   6.60  &   0.22  &   8.45  & RQ  \\
 823199  &  20.61  &   3.94  &   0.13  &  26.23  &   0.40  &   8.38  & RQ  \\
 823714  &  20.95  &   2.68  &   0.21  &  19.24  &   0.34  &   7.98  & RL  \\
 824176  &  21.48  &   6.19  &   0.19  &  25.54  &   2.31  &   8.77  & RL  \\
 824306  &  20.99  &   2.68  &   0.10  &  27.69  &   1.14  &   8.05  & RQ  \\
 824390  &  21.12  &   4.90  &   0.27  &  19.09  &   0.65  &   8.50  & RQ  \\
 824396  &  19.49  &   3.62  &   0.19  & 108.25  &   4.59  &   8.59  & RQ  \\
 824572  &  20.45  &   4.27  &   0.47  &  52.42  &   1.44  &   8.59  & RL  \\
 825363  &  21.70  &   4.54  &   0.23  &  10.70  &   0.36  &   8.32  & RQ  \\
 825899  &  20.85  &   3.50  &   0.92  &  32.38  &   1.96  &   8.32  & RQ  \\
 825906  &  22.05  &   6.09  &   0.10  &   6.52  &   0.24  &   8.47  & RQ  \\
 827274  &  20.82  &   2.89  &   0.05  &  17.00  &   0.37  &   8.02  & RQ  \\
 829667  &  19.29  &   5.60  &   0.27  & 218.38  &   3.40  &   9.12  & RQ  \\
 829682  &  21.60  &   4.97  &   0.04  &  12.63  &   0.21  &   8.43  & RQ  \\
 830510  &  22.17  &   6.76  &   0.23  &  17.24  &   0.85  &   8.76  & RQ  \\
 831077  &  20.58  &   4.13  &   0.11  &  55.50  &   3.90  &   8.57  & RQ  \\
 832354  &  20.12  &   4.45  &   0.09  &  67.77  &   1.49  &   8.68  & RQ  \\
 832715  &  20.15  &   7.90  &   0.14  &  50.76  &   0.81  &   9.12  & RL  \\
 832923  &  21.29  &   3.59  &   0.26  &  15.58  &   0.35  &   8.19  & RL  \\
 832961  &  21.86  &   3.33  &   0.24  &  12.29  &   0.71  &   8.08  & RQ  \\
 832963  &  22.40  &   4.93  &   1.25  &   6.90  &   0.55  &   8.30  & RQ  \\
 833273  &  21.94  &   5.23  &   0.38  &  14.22  &   1.83  &   8.50  & RQ  \\
 833541  &  21.11  &   4.99  &   0.30  &  10.49  &   0.34  &   8.40  & RQ  \\
 833712  &  22.03  &   4.76  &   0.22  &   2.43  &   0.44  &   8.06  & RQ  \\
 833817  &  20.57  &   5.56  &   0.39  &  27.18  &   0.45  &   8.69  & RQ  \\
 834079  &  21.41  &  11.18  &   0.53  &   7.70  &   0.31  &   9.03  & RL  \\
 834383  &  19.86  &   5.32  &   0.12  & 111.41  &   1.68  &   8.94  & RQ  \\
 834988  &  20.57  &   5.53  &   0.21  &  18.92  &   0.39  &   8.61  & RQ  \\
 835006  &  22.20  &   1.93  &   0.10  &   5.64  &   0.28  &   7.45  & RQ  \\
 835631  &  19.54  &   2.31  &   0.10  & 144.30  &   2.77  &   8.26  & RQ  \\
 835840  &  22.29  &   3.19  &   0.33  &   6.17  &   0.75  &   7.90  & RQ  \\
 836198  &  21.76  &   3.92  &   0.12  &  10.76  &   0.35  &   8.19  & RQ  \\
 836355  &  20.68  &   5.10  &   0.14  &  17.32  &   0.22  &   8.52  & RQ  \\
 837652  &  20.08  &   5.90  &   0.11  &  76.87  &   1.86  &   8.95  & RQ  \\
 837827  &  22.06  &   3.82  &   0.10  &   6.02  &   0.28  &   8.05  & RQ  \\
 837858  &  21.02  &   3.54  &   0.09  &  24.02  &   1.40  &   8.27  & RQ  \\
 838223  &  20.05  &   4.31  &   0.06  &  54.27  &   0.92  &   8.61  & RL  \\
 838610  &  21.88  &   3.94  &   1.10  &  12.12  &   1.13  &   8.22  & RQ  \\
 839188  &  20.55  &   4.83  &   0.07  &  28.03  &   0.36  &   8.57  & RL  \\
 839751  &  22.11  &   9.04  &   0.74  &   5.90  &   0.76  &   8.80  & RQ  \\
 841635  &  21.72  &   5.80\tablenotemark{f}  &   0.41  &   7.01  &   0.23  &   8.45  & RQ  \\
 843302  &  21.15  &   3.82  &   0.21  &  16.72  &   0.27  &   8.26  & RQ  \\
 843685  &  21.53  &   5.55  &   0.11  &   6.48  &   0.22  &   8.39  & RQ  \\
 844213  &  20.47  &   4.76  &   0.36  &  62.15  &   2.02  &   8.72  & RQ  \\
 845220  &  20.85  &   3.92  &   0.20  &  25.43  &   3.22  &   8.37  & RQ  \\
 845272  &  22.49  &   3.79  &   0.26  &   5.59  &   0.42  &   8.03  & RQ  \\
 845728  &  22.19  &   6.19  &   0.16  &   3.43  &   0.29  &   8.36  & RQ  \\
 845970  &  21.08  &   3.79  &   0.20  &  35.96  &   2.41  &   8.41  & RQ  \\
 846335  &  20.21  &   4.30\tablenotemark{f}  &   0.28  &  57.32  &   1.29  &   8.61  & RQ  \\
 847623  &  21.59  &   6.54  &   1.44  &   4.55  &   0.23  &   8.46  & RQ  \\
 848220  &  20.03  &   3.83  &   0.19  &  94.11  &   2.76  &   8.62  & RQ  \\
 851007  &  20.72  &   3.38  &   0.11  &  41.31  &   1.23  &   8.34  & RQ  \\
 900028  &  22.49  &   8.21  &   0.24  &   1.66  &   0.20  &   8.45  & RQ  \\
 900066  &  22.36  &   6.89  &   0.60  &   6.14  &   0.25  &   8.57  & RQ  \\
 950021  &  22.11  &   6.43  &   0.17  &  12.09  &   1.77  &   8.65  & RQ  \\

 \enddata
\tablenotetext{a}{Apparent $i$-band (AB) magnitude of the
  zCOSMOS spectroscopy target}   
\tablenotetext{b}{\MgII\ emission line FWHM in units of 1000 km
   s$^{-1}$}
 \tablenotetext{c}{Monochromatic continuum AGN luminosity in units of
   10$^{40}$ erg s$^{-1}$}
 \tablenotetext{d}{Logarithm of the black hole mass, as computed with
   Eq.(\ref{eq:virial}), adopting the MG08 calibration (see text for
   details)}
\tablenotetext{e}{RQ are objects detected in the VLA/COSMOS Survey at
  1.4 GHz \citep{bondi:08,schinnerer:07}}
\tablenotetext{f}{Objects with clear narrow absorption features in
  the \MgII\ line}
  \end{deluxetable}


 
 \end{document}